\documentclass[fleqn,usenatbib]{mnras}

\usepackage{newtxtext,newtxmath}
\usepackage[T1]{fontenc}

\DeclareRobustCommand{\VAN}[3]{#2}
\let\VANthebibliography\thebibliography
\def\thebibliography{\DeclareRobustCommand{\VAN}[3]{##3}\VANthebibliography}

\usepackage{graphicx}	% Including figure files
\usepackage{amsmath}	% Advanced maths commands
\usepackage{subfig}

\title[s- and r-process elements across the MW]{Origin of neutron capture elements with the Gaia-ESO survey: the evolution of s- and r-process elements across the Milky Way}

\author[M. Molero et al.]{
Marta Molero,$^{1,2}$\thanks{E-mail: marta.molero@phd.units.it}
Laura Magrini,$^{3}$
Francesca Matteucci,$^{1,2}$
Donatella Romano,$^{4}$
Marco Palla,$^{5}$
\newauthor
Gabriele Cescutti,$^{1,2}$
Carlos Viscasillas Vázquez,$^{6}$
Emanuele Spitoni$^{2,7}$
\\
$^{1}$Dipartimento di Fisica, Sezione di Astronomia, Università degli studi di Trieste, Via G.B. Tiepolo 11, I-34143 Trieste, Italy\\
$^{2}$INAF, Osservatorio Astronomico di Trieste, Via Tiepolo 11, I-34131 Trieste, Italy\\
$^{3}$INAF, Osservatorio Astrofisico di Arcetri, Largo E. Fermi 5, 50125 Firenze, Italy\\
$^{4}$INAF, Osservatorio di Astrofisica e Scienza dello Spazio, Via Gobetti 93/3, I-40129 Bologna, Italy\\
$^{5}$Sterrenkundig Observatorium, Ghent University, Krijgslaan 281 – S9, 9000 Gent, Belgium\\
$^{6}$Institute of Theoretical Physics and Astronomy, Vilnius University, Sauletekio av. 3, 10257 Vilnius, Lithuania\\
$^{7}$Université Côte d’Azur, Observatoire de la Côte d’Azur, CNRS, Laboratoire Lagrange, Bd de l’Observatoire, CS 34229, 06304 Nice cedex 4, France\\
}

\date{Accepted XXX. Received YYY; in original form ZZZ}

\pubyear{2022}

\begin{document}
\label{firstpage}
\pagerange{\pageref{firstpage}--\pageref{lastpage}}
\maketitle

\begin{abstract}
We investigate the origin of neutron-capture elements by analyzing their abundance patterns and radial gradients in the Galactic thin disc. We adopt a detailed two-infall chemical evolution model for the Milky Way, including state-of-the-art nucleosynthesis prescriptions for neutron capture elements. We consider r-process nucleosynthesis from merging neutron stars (MNS), magneto-rotational supernovae (MR-SNe) and s-process synthesis from low- and intermediate-mass stars (LIMS) and rotating massive stars. The predictions of our model are compared with data from the sixth data release of the \textit{Gaia}-ESO survey, from which we consider 62 open clusters with age $\mathrm{\gtrsim0.1\ Gyr}$ and $\mathrm{\sim1300}$ Milky Way disc field stars. We conclude that:  i) the [Eu/Fe] vs. [Fe/H] diagram is reproduced by both prompt and delayed sources, with the prompt source dominating Eu production; ii) rotation in massive stars significantly contributes to the first peak s-process elements, but MNS and MR-SNe are necessary to match the observations; iii) our model slightly underpredicts Mo and Nd, while accurately reproducing the [Pr/Fe] vs. [Fe/H] trend. Regarding the radial gradients, we find that: i) our predicted [Fe/H] gradient slope agrees with observations from \textit{Gaia}-ESO and other high-resolution spectroscopic surveys; ii) the predicted [Eu/H] radial gradient slope is steeper than the observed one, regardless of how quick the production of Eu is, prompting discussion on different Galaxy formation scenarios and stellar radial migration effects; iii) elements in the second s-process peak as well as Nd and Pr exhibit a plateau at low Galactocentric distances, likely due to enhanced enrichment from LIMS in the inner regions.
\end{abstract}


%%%%%%%%%%%%%%%%%%%%%%%%%%%%%%%%%%%%%%%%%%%%%%%%%%

%%%%%%%%%%%%%%%%% BODY OF PAPER %%%%%%%%%%%%%%%%%%

\section{Introduction}

The majority of elements beyond the Fe peak are produced by neutron capture processes which can be rapid (\textit{r-process}) or slow (\textit{s-process}) with respect to the $\beta$-decay of the nuclei. Understanding which are the astrophysical sites of these two processes has become one of the major challenges in stellar physics and chemical evolution. 

The r-process sites, in particular, are still under debate, with possible main producer candidates being supernovae (SNe) or merging of compact objects (neutron stars or neutron star-black hole, e.g. \citealp{Thielemann2017,Cowan2021}). Historically, the occurrence of the r-process has first been associated with core-collapse SNe (CC-SNe; \citealp{woosley1994}). However, during recent years it has been shown that hydrodynamical simulations of the neutrino-driven wind are not able to reach the extreme conditions necessary for the r-process (\citealp{arcones2013}). In addition to the neutrino-driven wind, a second mechanism has been proposed in order to explain observations of very energetic explosions. This mechanism relies on rapid rotation of the Fe core from which rotational energy can be extracted via a strong magnetic field. SNe explosions which may rely on that mechanism are named magneto-rotational SNe (MR-SNe; see e.g. \citealp{winteler2012,nishimura2015,Nishimura2017,reichert2021,reichert2023}). MR-SNe are rare with respect to normal CC-SNe, since the required rotation rates and high magnetic fields restrict the mechanism to a minority of progenitor stars (only 1\% of stars with initial mass larger than $\mathrm{10\ M_\odot}$, according to \citealp{2006woosley}). Despite their low rate, MR-SNe may be important contributors to the enrichment of galaxies with heavy elements (see \citealp{Arcones2023} for a recent review), especially at low metallicities. However, the dependence of their nucleosynthesis on the magnetic field is strong and only with high magnetic fields the r-process produces heavy elements up to the third r-process peak (\citealp{reichert2021}) and still more work is needed to get a much better picture (see also \citealp{2022Schatz}).

Thanks to the detection of the kilonova AT2017gfo associated with the gravitational wave event GW170817 (\citealp{Abbott2017}), MNS have been proven to be a primary site for the production of r-process material (\citealp{pian2017}). \citet{Watson2019} reported the identification of the neutron capture element Sr in the reanalysis of the spectra and \citet{Perego2022} computed the amount of Sr synthesized in the wind ejecta of MNS and found it to be consistent with the one required to explain the spectral features of AT2017gfo. However, although the observations point towards MNS as the major astrophysical r-process site, chemical evolution simulations still struggle to reproduce the abundance pattern of the [Eu/Fe] vs. [Fe/H] (with Eu being a typical r-process element) if MNS are the only producers of r-process material and realistic timescales for merging are assumed. In chemical evolution simulations, the timescales for the production of a certain chemical element are of great importance. For MNS the delay time for the coalescence is determined by the stellar nuclear lifetime and the delay due to gravitational radiation. Therefore, since the delay time can potentially assume a wide range of values, the timescales for the enrichment from MNS will depend on a delay time distribution (DTD, see \citealp{Simonetti2019, Greggio2021}). \citet{Matteucci2014} showed that the observed [Eu/Fe] abundance pattern in the Milky Way (MW) can be reproduced with only MNS as Eu producers if the timescale for the Eu pollution does not exceed $\mathrm{10-30\ Myr}$, which can be attained by assuming a gravitational delay of $\mathrm{1\ Myr}$ on top of the nuclear evolution timescale of the progenitors. If longer delays are assumed then another source of Eu should be included, especially at low metallicities. A large number of works (e.g. \citealp{Wehmeyer2015, Simonetti2019, Cote2019, Kobayashi2020, Cavallo2021, Molero2021}) point toward a scenario in which both a quick source and a delayed one produce r-process material. The delayed source is represented by MNS with a DTD while the quick source is usually represented by massive stars (e.g. MR-SNe). In such a way, the lack of Eu from MNS at low [Fe/H], due to longer delay times for merging, is compensated by the production from MR-SNe. However, if these sources are active also for the production of the r-process component of Ba, then the above models struggle to reproduce the observed [Ba/Fe] vs [Fe/H] trend at low metallicites (see \citealp{Molero2021_2}), suggesting that more investigation is needed.

Conversely, we have a better understanding of the formation sites for the s-process. The s-process can be decomposed into three sub-processes: i) the weak s-process, mainly responsible for the production of the first peak s-process elements (Y, Sr and Zr); ii) the main s-process which produces elements belonging to the second s-process peak (Ba, La and Ce) and iii) a strong s-process responsible for the production of the third s-process peak elements (e.g.: Pb; Au and Bi). Rotating massive stars are the main responsible for the production of the weak s-process. The source of neutrons is the reaction $\mathrm{^{22}Ne(\alpha,n)^{25}Mg}$ which takes place mainly during core He- and C-burning phases (\citealp{Pignatari2010,Frischknecht2016, Limongi2018}). The main s-process component occurs in low and intermediate mass stars (LIMS; $\mathrm{M<8\ M_\odot}$) during their asymptotic giant branch (AGB) phase, in which the neutron source is the reaction $\mathrm{^{13}C(\alpha,n)^{16}O}$ (\citealp{karakas2010,Fishlock2014,Cristallo2015}). Finally, low-metallicity low-mass AGB stars are responsible for the production of elements belonging to the third s-process peak through the strong s-process.

In this paper, we aim at improving our understanding of the origin of the neutron capture elements. For this purpose, we study both the abundance patterns and the radial gradients of five s-process (Y, Zr, Ba, La and Ce) and four mixed/r- process elements (Eu, Mo, Nd, and Pr) in the Galactic thin disc. We take advantage of the sixth data release of the \textit{Gaia}-ESO survey from which our sample has been collected. The sample consists of 62 open clusters (OCs) located between $\mathrm{\sim\ 5-20\ kpc}$ in Galactocentric distances, with ages from $\mathrm{0.1-7\ Gyr}$ and covering a metallicity range of $\mathrm{-0.5<[Fe/H]<0.4\ dex}$, together with $\mathrm{\sim 1300}$ MW disc field stars in the metallicity range of $\mathrm{-1.5<[Fe/H]<0.5\ dex}$ (see also \citealp{vanderswaelmen2023, Magrini2023}). We adopt a detailed two-infall chemical evolution model with an inside-out scenario of formation, extensively tested to explain the features observed in the MW disc (e.g. \citealp{Palla2020, spitoni2021}). With our model, we are able to follow the evolution of 40 different chemical species. 
 
 The paper is structured as follows. In Section \ref{sec: observational data} we present the \textit{Gaia}-ESO sample, in Section \ref{sec: the model} and \ref{sec: nucleosynthesis prescriptions} we describe the details of the chemical evolution model. Section \ref{sec: results} presents our results first for the [El/Fe] vs. [Fe/H] and then for the radial gradients of both [Fe/H] and the neutron capture elements. Finally, in Section \ref{sec: conlusion} we draw our summary and conclusions.

\section{Observational Data}
\label{sec: observational data}

The {\em Gaia}-ESO survey is a large public spectroscopic survey that observed for 340 nights at the VLT from the end of 2011 to 2018 using the FLAMES spectrograph \citep{randich22, gilmore22}. 
During the survey, FLAMES was used at intermediate spectral resolution with GIRAFFE, and at high resolution with UVES. 
In this work, we select the spectra of FGK stars obtained with UVES at R=47000, covering the spectral range 480.0$-$680.0~nm. 
The spectra were analysed by the {\em Gaia}-ESO Working Group (WG~11) dedicated to the analysis of FGK stars. We refer the reader to \citet{randich22, gilmore22} for a general description of the  structure of {\em Gaia}-ESO  and of the analysis procedure.   The final catalogue containing among others atmospheric parameters, elemental abundances, radial and projected rotational velocities is publicly available in the ESO archive \footnote{\url{https://www.eso.org/qi/catalogQuery/index/393}}. 
The high spectral resolution of UVES and the large  collecting area of VLT make it possible to obtain precise abundances of many neutron capture elements: the slow-process elements Y, Zr, Ba, La, and Ce, and the mixed/r-process elements Mo, Pr, Nd, and Eu. Throughout the paper, we  use these abundances, normalizing them to the Solar scale as in \citet{viscasillas22, Magrini2023}. 

Other surveys as the Apache Point Observatory
Galactic Evolution Experiment \citep[APOGEE-1 and APOGEE-2;][]{apogee} and  the Galactic Archaeology with HERMES \citep[GALAH;][]{galah} are providing
abundances of some neutron-capture elements as well. 
For example, APOGEE provides for a limited percentage of stars abundances of Ge, Rb, Ce, Nd, and Yb, but only Ce has been used in scientific works \citep{cunha17, donor20}. Recently, \citet{hayes22} explored weak and blended species  in the APOGEE database, providing also new and improved abundances of Ce and Nd. The GALAH survey provided in its third data release \citep{galahdr3} abundances of several neutron capture elements, namely Y, Ba, La, Rb, Mo, Ru, Nd, Sm, and of Eu. However, their accuracy is often limited because of the lower spectral resolution.  The quality of the neutron-capture abundances in the UVES spectra of {\em Gaia}-ESO remains, thus,  unrivalled. 

\subsection{The data samples}
In this work, we used two different samples of stars: a first sample composed by stars that are members of open clusters, and a second sample of field stars. 
For the former, we benefit of the large sample of stars members of open star clusters, located at various Galactocentric distances and covering a wide age range, from a few million years to about 7 Gyr.  Star clusters, containing groups of coeval and chemically homogeneous stars, allow, indeed, a more accurate determination of age and chemical properties with respect to field stars. 
The latter is, instead, composed of stars of the main sequence turn off (MSTO), 
%located in the solar neighborhood 
\citep[see][for the description of the selection function]{Stonkute16}. 

\subsubsection{The sample of open clusters}
In the present paper, we consider among the sample of open clusters observed by {\em Gaia}-ESO, the 62  clusters older than 100~Myr, as done in \citet{Magrini2023}. 
The motivation is twofold: younger stars might be affected by problems in the spectral analysis, as shown by  \citep{baratella20, baratella21, spina21}; young clusters represent only the last moments of global galactic chemical evolution, with negligible variations of the abundance with respect to the overall time scale. The distribution in age and distances of our sample open clusters is given in \citet[see their Fig.~1]{viscasillas22}. 
For each cluster, we considered the average abundances of its member stars. The membership analysis is performed as in \citet{viscasillas22}, based on three-dimensional kinematics, complementing the radial velocities  from {\em Gaia}-ESO with proper motions and parallaxes from {\em Gaia} {\sc dr3} \citep{gaiadr3}.
Ages and Galactocentric distances are homogeneously derived with {\em Gaia} {\sc dr2} data in \citet{cantat20}. 
In  the paper, we use the open cluster sample to trace the abundance radial gradients, and thanks to the wide age range, also their evolution over time.  

\subsubsection{The sample of field stars}

The sample of field stars is composed, as in \citet{viscasillas22}, by stars selected through the GES\_FLD keywords related to the field stars (GES\_MW for general Milky Way fields, GES\_MW\_BL for fields in the direction of the Galactic bulge, GES\_K2 for stars observed in Kepler2 (K2) fields, GES\_CR for stars observed in CoRoT fields, and benchmark stars GES\_SD), and stars which are non-members of open clusters. 
We combined the two samples, applying a further selection on the signal-to-noise ratio (SNR) and on the uncertainties on the stellar parameters: SNR > 20; $\sigma T_{\rm eff} < 150$~K, $\sigma{\rm log}\,g < 0.25$, $\sigma {\rm [Fe/ H]}< 0.20$ and $\sigma \xi <$ 0.20 $km~ s^ {-1}$. A final selection was introduced considering only stars with at least one measurement of the abundances of one of the considered neutron capture elements, and with an uncertainty $e A(El) < 0.1$. These selections produce a sample of approximately 1300 stars. 
Due to the selection function of the {\em Gaia}-ESO survey \citep[see][]{Stonkute16}, this sample is dominated by stars at the MSTO, with some giant stars which are non members of open clusters.  
Due to the wide metallicity range covered by the field stars, we use them to study the evolution in the [El/Fe] vs [Fe/H] planes.

\section{The Model}
\label{sec: the model}

The adopted chemical evolution model derives from the two-infall model originally developed by \citet{Chiappini1997}. Here we use the revised version of \citet{Palla2020} focusing our study on the thick and thin discs only, without taking into account the evolution of the Galactic halo. 

The two-infall model assumes that the Galaxy forms as a result of two infall episodes. The first one forms the thick disc while the second one, delayed with respect to the first, is responsible for the creation of the thin disc. The composition of the infalling gas is assumed to be primordial. The disc is approximated by independent rings  2 kpc wide.

The basic equations which describe the evolution of the fraction of gas mass in the form of a generic element $i$, $G_{i}$, are (see \citealp{Matteucci2012}):

\begin{equation}
    \dot{G}_i(R,t)=-\Psi(R,t)X_i(R,t)+R_i(R,t)+\dot{G}_{i,inf}(R,t).
\end{equation}

The first term on the right-hand side of the equation represents the rate at which the chemical elements are subtracted to the ISM to be included in stars. $X_i\mathrm{(R,t)}$ represents the abundance by mass of a given elements $i$, while $\Psi\mathrm{(R,t)}$ is the star formation rate, here parametrized according to the Schmidt-Kennicutt law (\citealp{Kennicutt1998}):

\begin{equation}
    \Psi(R,t)\propto\nu\sigma_{gas}(R,t)^k,
\label{eq: equation of chemical evolution}
\end{equation}
\\
where $\sigma_\mathrm{{gas}}$ is the surface gas density, $\mathrm{k=1.5}$ is the law index and $\nu$ is the star formation efficiency expressed in $\mathrm{Gyr^{-1}}$. This last term is assumed to vary with the Galactocentric distance as in \citet{Palla2020}.

The second term, $R_i\mathrm{(R,t)}$, represents the fraction of matter that is restored to the ISM in the form of the element $i$ through stellar winds, SN explosions, novae and MNS. Namely, it represents the rate at which chemical elements are restored to the ISM by all stars dying at the time $t$. It depends also on the initial mass function (IMF), which here is the \citet{Kroupa1993}'s one.

The last term in equation \ref{eq: equation of chemical evolution} is the gas accretion rate. In the two-infall model, it is computed in the following way:

\begin{equation}
    \dot{G}_{i,inf}(R,t)=A(R)X_{i,inf}e^{-t/\tau_1}+    
    \theta(t-t_{max})B(R)X_{i,inf}e^{-(t-t_{max})/\tau_2},
\end{equation}
\\
where $G_{i,\mathrm{inf}}\mathrm{(R,t)}$ is the infalling material in the form of the elements $i$ and $X_{i,\mathrm{inf}}$ is the composition of the infalling gas, here assumed to be primordial. $\mathrm{\tau_1}$ and $\mathrm{\tau_2(R)}$ are the infall time-scales for the thick and thin discs, respectively. We fix $\mathrm{\tau_1=1\ Gyr}$ and let $\mathrm{\tau_2}$ vary with the radius according to the inside-out scenario (e.g.: \citealp{Matteucci1989, Romano2000, Chiappini2001}) as:

\begin{equation}
    \tau_2(R)=\Big(1.033\frac{R}{kpc}-1.267\Big)\ Gyr.
\end{equation}

$t_\mathrm{max}$ is the time for the maximum infall onto the thin disc and it corresponds to the end of the thick disc phase and the start of the second infall episode. The typical value assumed for $t_\mathrm{max}$ in previous models (e.g.: \citealp{Chiappini2001, 2009Spitoni, Romano2010, Grisoni2018}) is $\mathrm{\sim 1\ Gyr}$. However, more recent works found out that the gap between the formation of the two discs should be higher in order to reproduce both stellar abundance constraints and ages. Here, we follow the prescriptions adopted in \citet{Palla2020}, who found a best value of $t_{\mathrm{max}} \mathrm{\simeq 3.25\ Gyr}$ (in agreement with \citealp{Spitoni2019, Spitoni2020}). 

The parameters $A(R)$ and $B(R)$ are fixed in order to reproduce the present time total surface mass densities of the thick and thin discs, as a function of the radius. Here we assume that the surface mass densities of the discs both follow exponential laws. In particular we adopt the following profiles for the thin and thick disc, respectively:

\begin{equation}
    \Sigma_{thin}(R)=\Sigma_{0,thin}e^{-R/3.5},
\end{equation}

\begin{equation}
    \Sigma_{thick}(R)=\Sigma_{0,thick}e^{-R/2.3}.
\end{equation}
\\
Where $\Sigma_{0,\mathrm{thin}}\mathrm{=531\ M_\odot\ pc^{-2}}$ is the central surface mass density,
%$\mathrm{R_d=3.5\ kpc}$ is the thin disc scale length
and $\Sigma_{0,\mathrm{thick}}$ is fixed in order to obtain $\Sigma_{\mathrm{thick}}\mathrm{(8\ kpc)=12\ M_\odot\ pc^{-2}}$. As explained in \citet{Palla2020}, these choices for the discs surface mass densities allow us to obtain $\Sigma_{\mathrm{thin}}\mathrm{(8\ kpc)\sim54\ M_\odot pc^{-2}}$ (in agreement with \citealp{bovyrix2013} and \citealp{read2014}) and a ratio $\Sigma_{\mathrm{thin}}\mathrm{(8\ kpc)}/\Sigma_{\mathrm{thick}}\mathrm{(8\ kpc)\sim4}$ (in agreement with \citealp{Spitoni2020}).

\section{Nucleosynthesis Prescriptions}
\label{sec: nucleosynthesis prescriptions}

For all stars sufficiently massive to die in a Hubble time, the following stellar yields have been adopted:

\begin{itemize}
    \item For LIMS ($\mathrm{1 \leq M/M_\odot \leq 6}$) we adopted the non-rotational set of yields available on the web pages of the FRUITY data base\footnote{http://fruity.oa-teramo.inaf.it} (\citealp{Cristallo2009, Cristallo2011, Cristallo2015}).
    \item For massive stars we implemented \citet{Limongi2018}'s recommended yield set R where mass loss and rotation are taken into account.
    \item For Type Ia SNe (SNeIa) we assumed the single-degenerate scenario (see \citealp{Matteucci2001, Palla2021} for details) for the progenitors, in which SNe arise from the explosion via C-deflagration of a CO white dwarf in a close binary system when it has almost reached the Chandrasekhar mass due to accretion from its red giant companion. The adopted stellar yields are from \citet{Iwamoto1999} (model W7).
    \item We consider also chemical enrichment from novae. They do not affect the heavy elements treated here, but they can be important for the production of $\mathrm{^{7}Li}$ and CNO isotopes (see \citealp{2007josè}).
\end{itemize}

\subsection{Heavy elements production}

All the neutron capture elements studied in this work (apart from Eu) are assumed to be partially produced by the r- and s-processes.

The s-process nucleosynthesis takes place in LIMS during the asymptotic giant branch (AGB) phase and in rotating massive stars, with yields specified in the previous paragraph. Neutrons are produced via the reactions $\mathrm{^{13}C(\alpha,n)^{16}O}$ and $\mathrm{^{22}Ne(\alpha,n)^{25}Mg}$, with the former reaction being the dominant contribution in low-mass stars and the latter in more massive AGBs (see \citealp{Cristallo2011,Cristallo2015} for details) and massive stars (\citealp{2012longland}). 

For the r-process nucleosynthesis we considered two channels: MNS and MR-SNe. MNS are computed as systems of two neutron stars of $\mathrm{1.4\ M_\odot}$ with progenitors in the $\mathrm{9-50\ M_\odot}$ mass range. Their rate is computed as the convolution between a given delay time distribution (DTD) and the SFR, as:
\begin{equation}
 R_{MNS}(t)=k_{\alpha} \int_{\tau_{i}}^{min(t,\tau_{x})} \! \alpha_{MNS}(\tau)\Psi(t-\tau)f_{MNS}(\tau) \, \mathrm{d}\tau.
\label{eq:ratemns}
\end{equation}
$k_\alpha$ is the number of neutron star progenitors per unit mass in a stellar generation and the $\mathrm{\alpha_{MNS}}$ parameter is the fraction of stars in the correct mass range which can give rise to a double neutron star merging event. In principle, $\mathrm{\alpha_{MNS}}$ can vary with time, but here it is assumed to be constant and it is fine tuned in order to reproduce the latest estimation of the MNS rate of \citet{Abbott2021}. The DTD represents the probability that the merging event would happen at a certain time $t$ after the formation of the progenitor binary system. For MNS, the delay between the creation of the progenitors binary system and the merging event is the sum between the nuclear lifetime of the secondary component and the delay due to the gravitational radiation. The adopted DTD in this work is the following (for a detailed discussion see \citealp{Simonetti2019, Greggio2021}): 

\begin{equation}
    f_{MNS}(\tau)\propto 
    \begin{cases}
    \ 0 \qquad \text{if} \qquad \tau < 10\ \mathrm{Myr} \\
    \ p_1 \qquad \text{if} \qquad  10 < \tau/\mathrm{Myr} < 40  \\
    \ p_2\tau^{0.25\beta -0.75}(M_{m}^{0.75(\beta + 2.33)}-M_{M}^{0.75(\beta + 2.33)}) \\ 
    \ \text{if} \qquad  40\ \mathrm{Myr} < \tau < 13.7\ \mathrm{Gyr}
    \end{cases}
\label{eq: DTD}
\end{equation}

with $\beta=-0.9$, $M_m$ and $M_M$ being the minimum and maximum total mass of the binary system and $p_1$ and $p_2$ are chosen in order to obtain a continuous and normalized function. 

The yields from MNS of the various elements considered in this study have been obtained as in \citet{Molero2021_2} by assuming a scaling relation between them and those of Sr. The adopted yield of Sr is equal to $\mathrm{Y_{Sr}^{MNS}=1\times10^{-4}\ M_\odot}$, which corresponds to that measured by \citet{Watson2019} in the reanalysis of the spectra of the kilonova AT2017gfo multiplied by a factor of 10 (see \citealp{Molero2021_2} for details). 

Although MR-SNe are theorized to be among the most important contributors to the enrichment of r-process material, \citet{2006woosley} speculated that only 1\% of stars with initial mass $\mathrm{\ge 10\ M_\odot}$ can have the necessary conditions to die as a MR-SNe. A common assumption in chemical evolution models is that only 10\% of all stars with initial mass in the $\mathrm{10-80\ M_\odot}$ range end their lives as MR-SNe (e.g. \citealp{cescutti2014,cescutti2015,rizzuti2019,Rizzuti2021,Molero2021_2}). Both the percentage of stars, their mass range and the yields of r-process material are free parameters in chemical evolution simulations and they are usually fine tuned in order to reproduce the observations of abundances. In order to avoid degeneracy issues, in this work we fixed the mass range and yields, while keeping the percentage of stars as a free parameter. The mass range is reduced to $\mathrm{10-25\ M_\odot}$ in order to be consistent with the set of yields adopted for massive stars. In fact, in \citet{Limongi2018}'s set R stars more massive than $\mathrm{25\ M_\odot}$ are assumed to fully collapse to a black hole and their chemical enrichment is due to the stellar winds. Therefore, in these conditions it would be impossible for the star to develop magnetic fields strong enough to generate a MR-SNe. It must be noted that this variation in the mass range will not produce a too significant difference in the results, because of the adopted IMF (\citealp{Kroupa1993}), which is known to be top-light, i.e. which disfavours the presence of very massive stars due to its steep high mass end slope. The set of yields adopted for the MR-SNe is the one of \citet{Nishimura2017}, their model L0.75, chosen in order to be consistent with the best model of our previous work (\citealp{Molero2021_2}). For some of the studied elements for which those yields predicted a much higher/lower production (Y, Zr, Mo, Nd and Pr) they have been scaled to the one of Eu according to the solar abundances. Finally, the percentage of stars able to explode as MR-SNe has been fine tuned in order to fit the observed [Eu/Fe] vs. [Fe/H] relation (see next sections), and set to 20\%.

In Table \ref{tab: nucleosynthesis prescriptions} we summarize the adopted nucleosynthesis prescriptions.

\begin{table}
\centering
\begin{tabular}{lcccc}
\hline
   Model &  $\mathrm{v_{MS}\ (km/s)}$  &   LIMS  &   MR-SNe   & MNS \\
\hline
R-0 & 0 & \checkmark & \checkmark & \checkmark \\
R-150 & 150 & \checkmark & \checkmark & \checkmark \\
R-300 & 300 & \checkmark & \checkmark & \checkmark \\
\hline
R-150 MNS & 150 & \checkmark & X & \checkmark \\
R-150 MRD & 150 & \checkmark & \checkmark & X \\
\hline
noR-0 & 0 & \checkmark & X & X \\
noR-150 & 150 & \checkmark & X & X \\
noR-300 & 300 & \checkmark & X & X \\
\hline
\end{tabular}%
\caption{\label{tab: nucleosynthesis prescriptions}Nucleosynthesis prescriptions. In the $\mathrm{1^{st}}$ column we report the name of the model, in the $\mathrm{2^{nd}}$ the initial rotational velocities for massive stars. In the $\mathrm{3^{rd}}$, $\mathrm{4^{th}}$ and $\mathrm{5^{th}}$ columns we list whether LIMS, MR-SNe and MNS channels are active or not, respectively. We point out that in the case of model 'R-150 MNS' MNS are assumed to merge with a short and constant time delay of $\mathrm{10\ Myr}$ instead that with a DTD.}
\end{table}

%\begin{figure}
%    \centering
%    \includegraphics[width=1\columnwidth]{Images/SFR.png}
%    \caption{Time evolution of the SFR as predicted by our models at various Galactocentric distances. Right corner plot: predicted SFR in the solar neighborhood compared to the present day estimates by \citet{Guesten1982} and \citet{Prantzos2018}.}
%    \label{fig: SFR}
%\end{figure}

%\begin{figure}
%    \centering
%    \includegraphics[width=1\columnwidth]{Images/rate.png}
%    \caption{Predicted SNeIa, SNeII and MNS rates compared to present day observations from \citet{cappellaro1999} (for SNe) and estimate from \citet{Abbott2021} (for MNS).}
%    \label{fig: Rates}
%\end{figure}

\begin{figure}
    \centering
    \includegraphics[width=1\columnwidth]{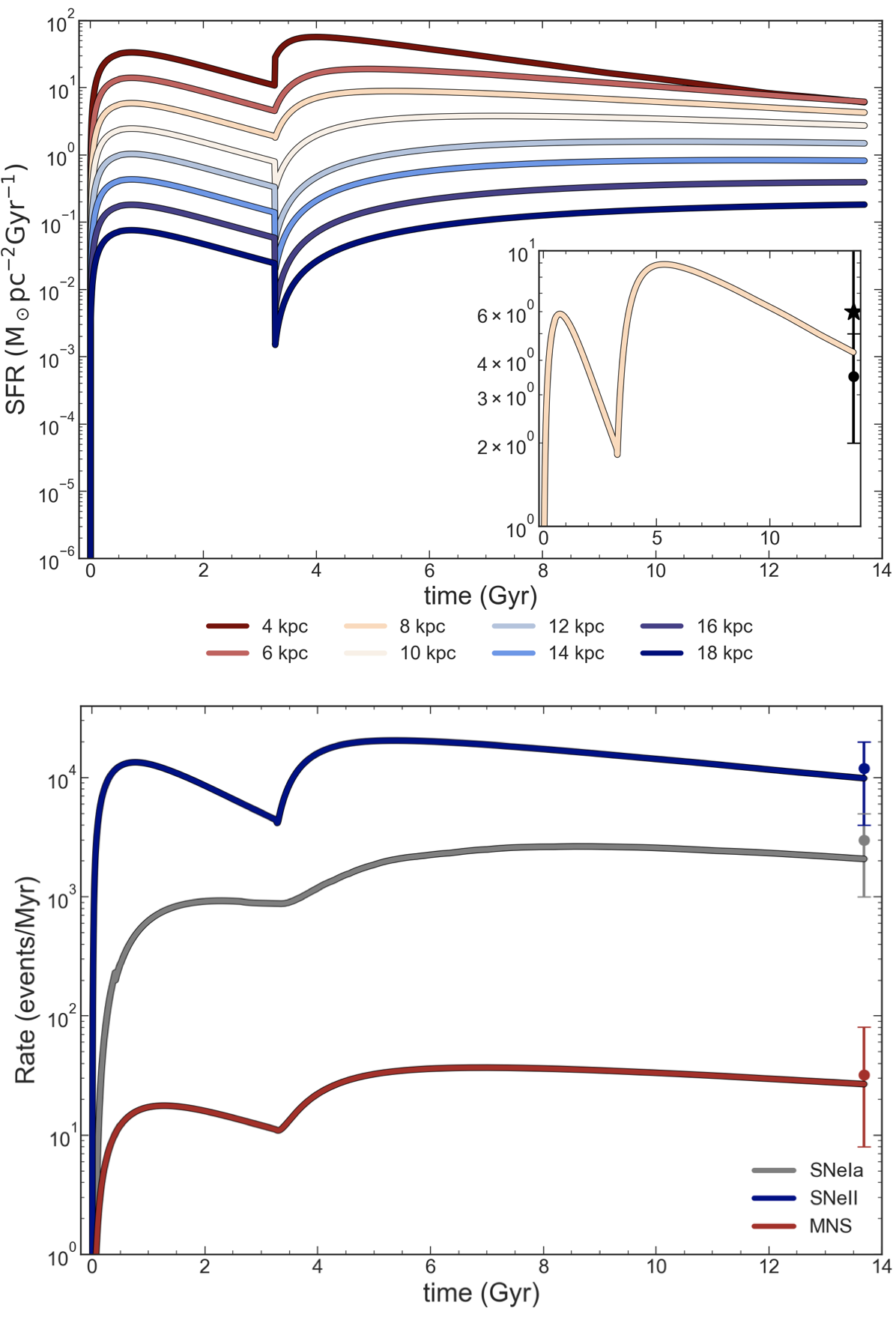}
    \caption{Upper panel: time evolution of the SFR as predicted by our models at various Galactocentric distances. Right corner plot: predicted SFR in the solar neighborhood compared to present day estimates (\citealp{Guesten1982,Prantzos2018}); lower panel: predicted SNeIa, SNeII and MNS rates compared to present day observations from \citet{cappellaro1999} (for SNe) and estimate from \citet{Abbott2021} (for MNS).}
    \label{fig:rate}
\end{figure}

\section{Results}
\label{sec: results}

Before discussing the comparison between our model predictions and the relevant observations for neutron capture elements, we show the evolution of some important quantities as functions of time. 

In the upper panel of Figure \ref{fig:rate}, we report the time evolution of the SFR as predicted by our model at different Galactocentric radii. In contrast with \citet{Grisoni2018}, the SFR during the thick disc phase is not the same for every Galactocentric distance up to $\mathrm{18\ kpc}$, since we do not assume a constant thick disc mass density, but rather an exponentially decaying surface density profile, as described in \citet{Palla2020}. As in the previously mentioned works, even without assuming a threshold for the SF, we are still able to obtain a quenching in the SF between the thick and the thin disc phases. The observed SFR in the solar neighborhood (see \citealp{Guesten1982,Prantzos2018}) is well reproduced by our model, as shown in the zoomed plot.

In the lower panel of Figure \ref{fig:rate}, we report our predictions for the rates of SNeIa, SNeII and MNS, averaged over the whole disc. The observational data are taken from \citet{cappellaro1999} for SNeIa and SNeII. For MNS we consider the latest cosmic rate observed by \citet{Abbott2021}, i.e., $\mathrm{320^{+490}_{-240}\ Gpc^{-3}yr^{-1}}$. We then applied the same conversion procedure developed by \citet{Simonetti2019} in order to convert the cosmic rate into a Galactic one. The rate so obtained is $\mathrm{R_{MNS}=32^{+49}_{-24}\ Myr^{-1}}$, in agreement within the error bars with the rate of \citet{kalogera2004}, derived from binary pulsars.

\subsection{Abundance ratios vs. metallicity trends}

In the next sections we will show results for the evolution of the [El/Fe] vs. [Fe/H] abundance patterns of the neutron capture elements studied in this work. First, we will discuss results for Eu, a pure r-process element, and then for the other s-process and mixed/r- process elements.

\begin{figure*}
    \centering
    \includegraphics[width=1\textwidth]{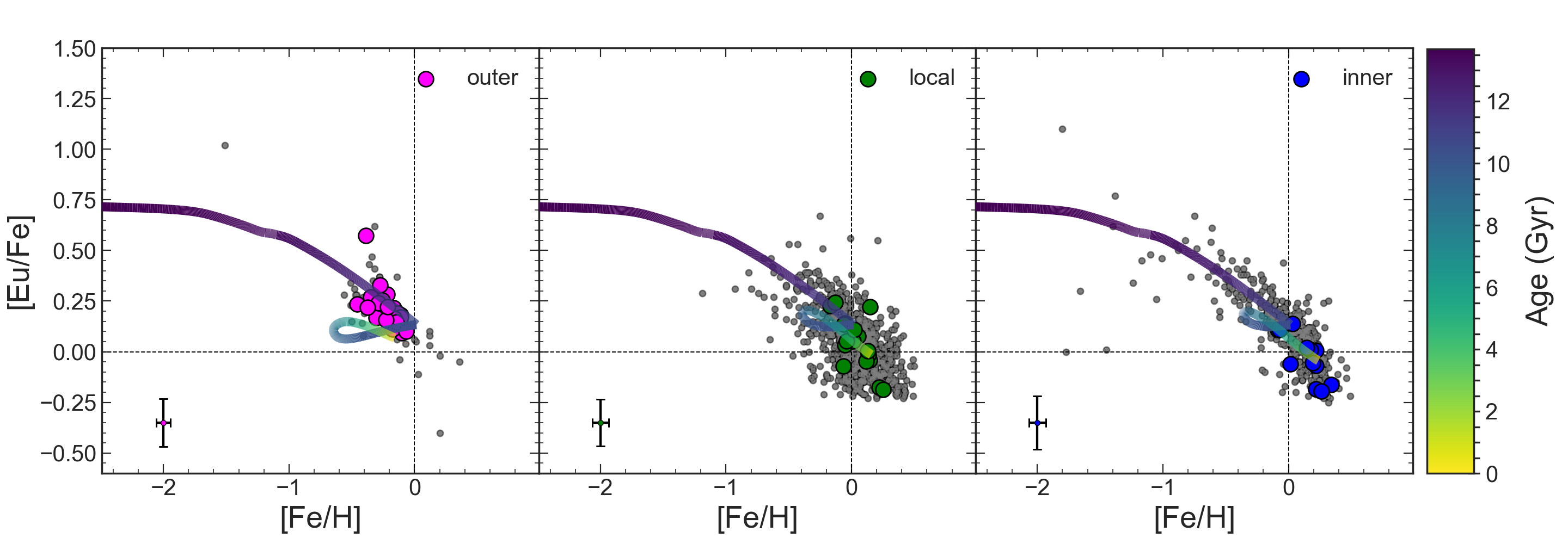}
    \caption{Predicted [Eu/Fe] vs [Fe/H] abundance patterns for the outer ($\mathrm{R_{GC}=12\ kpc}$), local ($\mathrm{R_{GC}=8\ kpc}$) and inner disc ($\mathrm{R_{GC}=6\ kpc}$). The curves refer to the predictions of model R-150 (see Table \ref{tab: nucleosynthesis prescriptions})and are color coded by the age of the stars created by the chemical evolution code. The grey small dots refer to the sample of field stars.}
    \label{fig: EuFe}
\end{figure*}

\subsubsection{Europium}

In Figure \ref{fig: EuFe} we report the [Eu/Fe] vs. [Fe/H] abundance pattern traced by the observational data (field stars and OCs) and compare it to the predictions of our models. As reported by \citet{vanderswaelmen2023}, who investigated the same Eu data, only for metallicity lower than $\mathrm{\sim -0.8\ dex}$ it is possible to distinguish the famous Eu plateau at [Eu/Fe] $\mathrm{\sim 0.4\ dex}$ for the field stars sample. Moreover, the plateau is visible more clearly only for stars belonging to the inner disc ($\mathrm{R_{GC}<7\ kpc}$). For inner disc stars, a scatter of $\mathrm{\sim 1\ dex}$ is present below [Fe/H] $\mathrm{\sim -1.8}$. However, this is due most probably to measurement errors rather than the stochastic enrichment of Eu characteristic of the halo (see \citealp{cescutti2014, cescutti2015, wanajo2021}). On the other hand, the field stars sample does not extend below $\mathrm{[Fe/H]=-1.2}$ ($\mathrm{-0.5}$) $\mathrm{dex}$ for $\mathrm{7<R_{GC}/kpc<9}$ ($\mathrm{R_{GC}>9\ kpc}$). The OC sample well overlaps with the field one for all $\mathrm{R_{GC}}$ and it is affected by a lower scatter, especially in the inner disc region.

Figure \ref{fig: EuFe} shows results of model R-150 computed at $\mathrm{R_{GC}=12\ kpc}$, $\mathrm{R_{GC}=8\ kpc}$ and $\mathrm{R_{GC}=6\ kpc}$ compared with outer-disc, local and inner-disc data, respectively. No difference is expected if the rotational velocity of massive stars is changed, since in our model rotation does not affect Eu production. The curves are color coded by the age of the stars created by the chemical evolution code. At $\mathrm{Age\simeq10.44\ Gyr}$, we notice the characteristic loop feature of the model due to the second infall phase. The accretion of pristine gas has the effect of decreasing the metallicity while having little impact on the [Eu/Fe] ratio. When the SF resumes, a rise in the [Eu/Fe] ratio is produced, then the ratio decreases while [Fe/H] increases because of the enrichment from SNeIa (see also \citealp{Spitoni2019,Spitoni2020, Palla2021}). The model tends to overestimate the age of the clusters in the outer zone, since the OCs have Age $\mathrm{\lesssim 7\ Gyr}$. However, the observed [Eu/Fe] vs. [Fe/H] trend is overall well reproduced for all the three different regions. We fixed the percentage of MR-SNe in order to fit the main trend, rather than the solar value. In order to reproduce the solar abundance of Eu, a smaller percentage of MR-SNe progenitors should be assumed, of the order of $\mathrm{15\%}$. The local and the inner-disc curves slightly underestimate the metallicity reached by the observational data. This may be due to a too low SF efficiency and/or fraction of Type Ia SN systems (SNeIa). However, the model seems to reproduce rather well the SF and SNIa rates at the present time in the solar neighborhood (see Figure \ref{fig:rate}). We also remind that in our model we do not include stellar migration effects which in principle can help reproducing the stars with larger [Fe/H] values (see e.g. \citealp{spitoni2015, palla2022}). We confirm once again that the best scenario is the one in which both a quick source and a delayed one are responsible for the production of Eu. Here, the quick source is represented by MR-SNe and the delayed one by MNS with a DTD. This is not a novelty in chemical evolution simulations (e.g.: \citealp{Matteucci2014, cescutti2015, Simonetti2019, Cote2019, Molero2021_2}). However, it must be noted that the quick source completely dominate the production of Eu. In fact, without the contribution from MNS, we would still be able to reproduce the observed abundance pattern. On the other hand, the [Eu/Fe] vs. [Fe/H] pattern cannot be reproduced if MNS were the only producers of Eu, if a DTD is adopted. Only by assuming a constant and short time delay ($\mathrm{\sim 1\ Myr}$) it is possible to explain the Eu enrichment as due to MNS alone (see \citealp{Matteucci2014}). Our model and prescriptions are slightly different from those adopted in \citet{vanderswaelmen2023}, but we confirm their same conclusion: Eu is produced mainly by a quick source and there is no need for an additional source at late times, at least in order to reproduce the observed Eu abundance pattern in the thin disc. However, it must be pointed out that MNS are the only source of heavy elements observed up to date, and because of that they cannot be excluded from chemical evolution simulations.

\begin{figure}
    \centering
    \includegraphics[width=1\columnwidth]{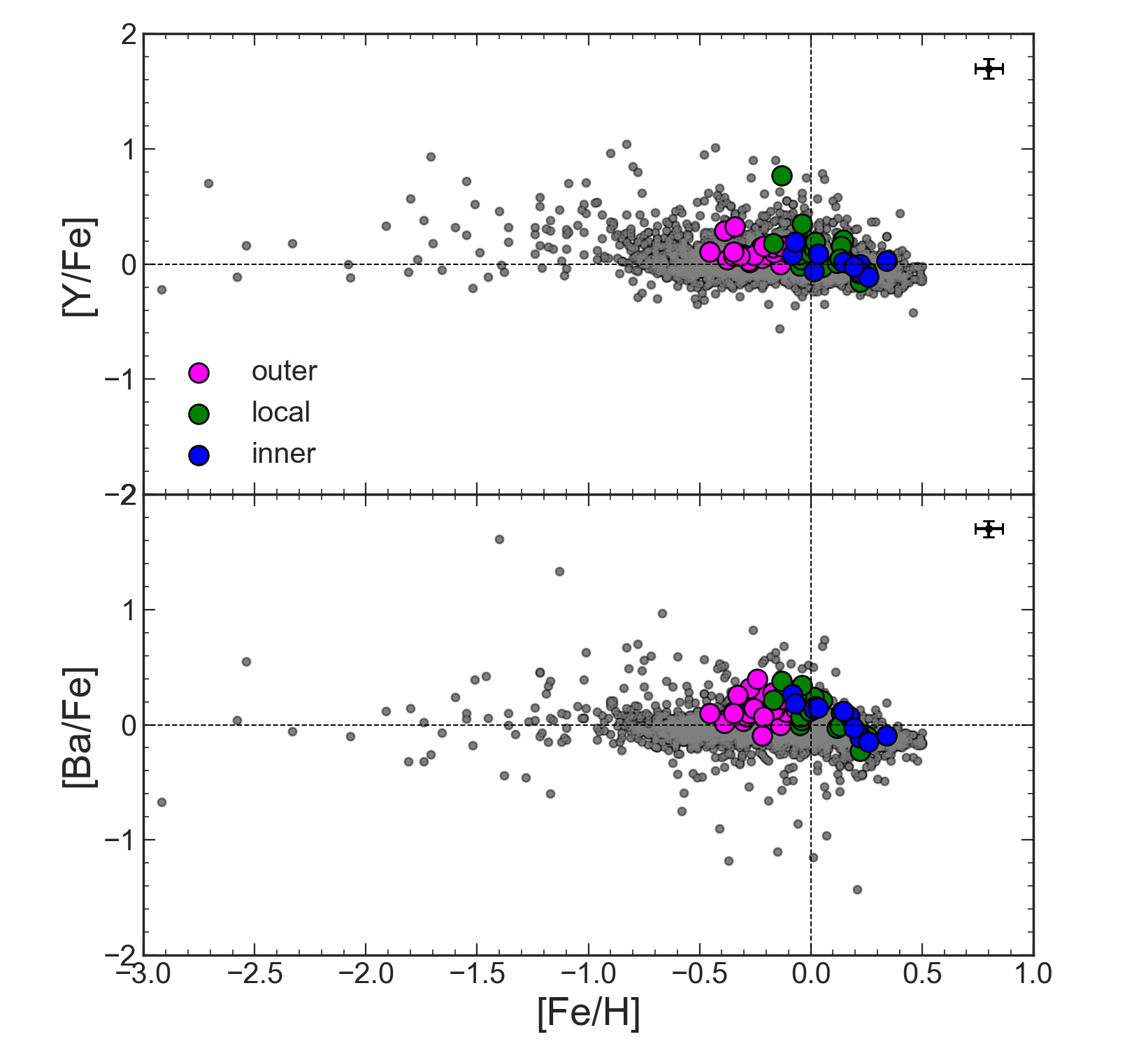}
    \caption{[Y/Fe] and [Ba/Fe] vs. [Fe/H] for our sample of field stars (in grey) and OCs (magenta, green and blue dots) at all Galactocentric distances.}
    \label{fig: Y and Ba obs only}
\end{figure}

\begin{figure*}
    \begin{center}
    \subfloat{\includegraphics[width=1\textwidth]{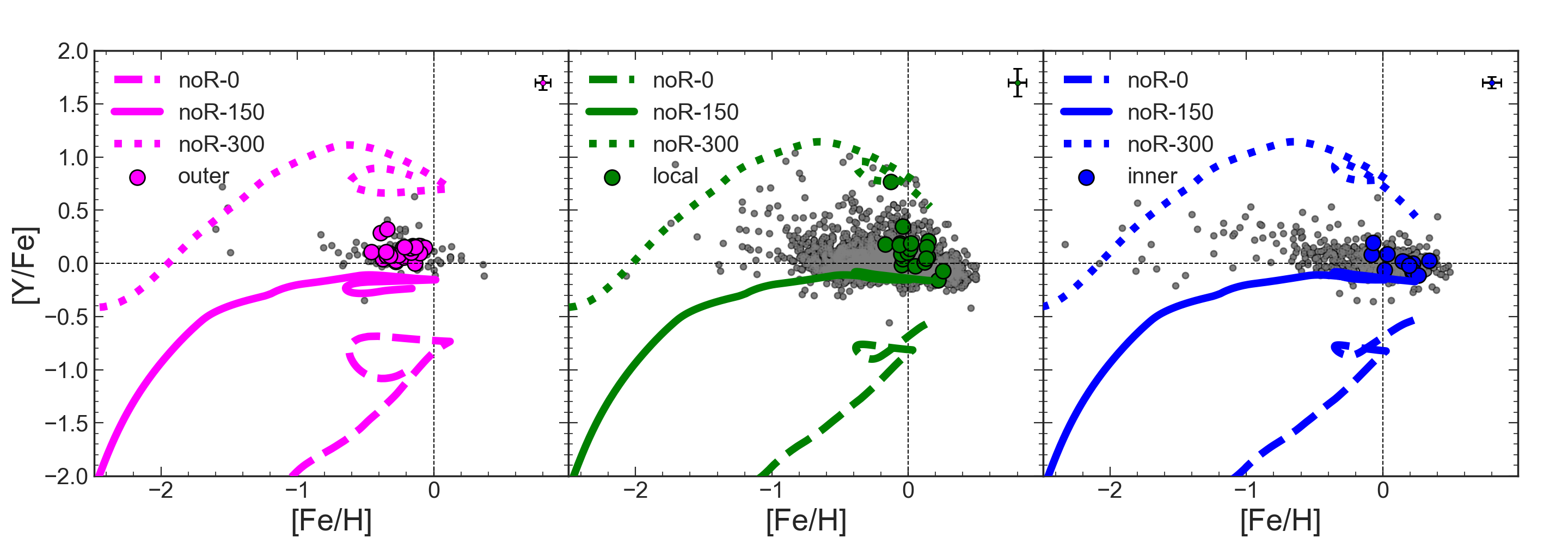}\label{fig:a}}
    \vspace{-0.4cm}
    \subfloat{\includegraphics[width=1\textwidth]{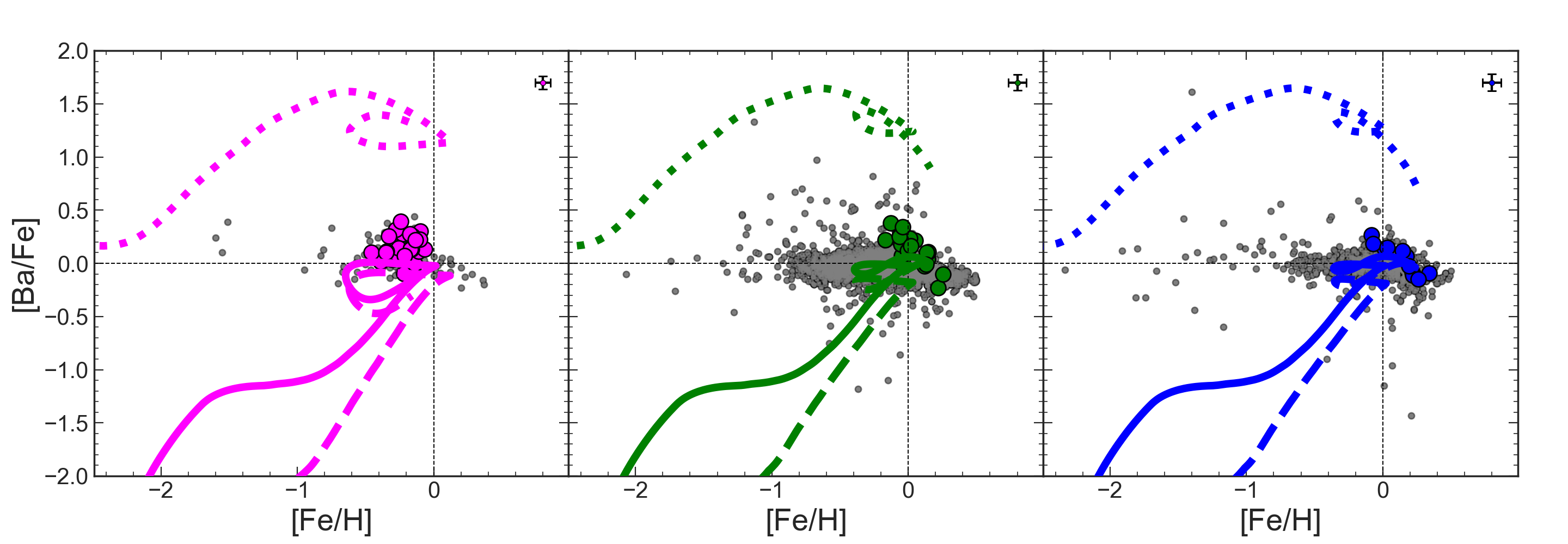}\label{fig:b}}%
    \caption{Predicted abundance patterns for [Y/Fe] and [Ba/Fe] vs. [Fe/H] for outer ($\mathrm{R_{GC}=12\ kpc}$), local ($\mathrm{R_{GC}=8\ kpc}$) and inner ($\mathrm{R_{GC}=6\ kpc}$) disc regions. We assume that only massive stars and LIMS are Y and Ba producers. The three lines in each plot corresponds to different initial rotational velocities of massive stars (see legend and Table \ref{tab: nucleosynthesis prescriptions}).}%
    \label{fig: s-elements solar RMS+reduced LIMS}%
    \end{center}
\end{figure*}

\begin{figure*}
    \centering
    \includegraphics[width=1\textwidth]{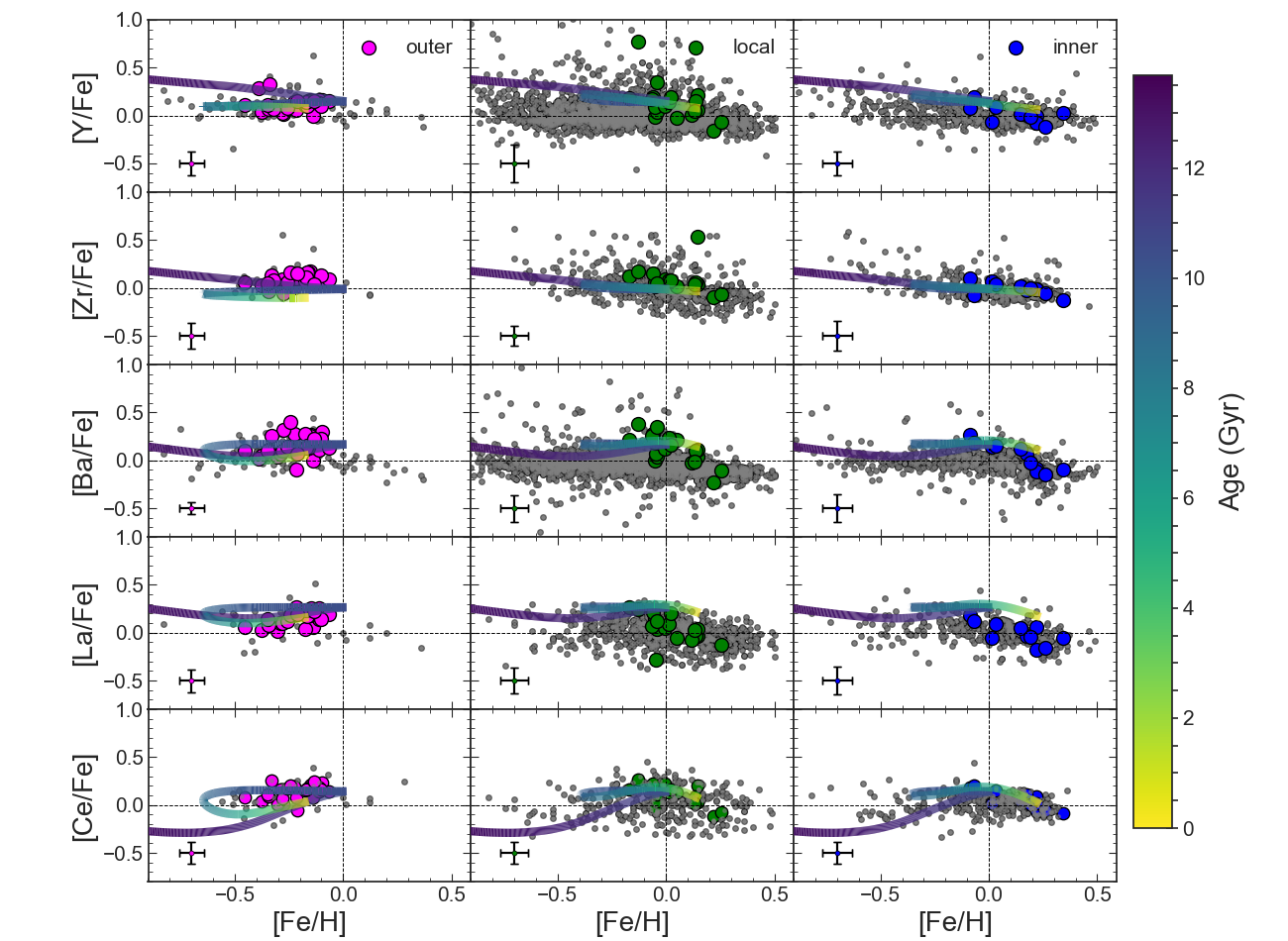}
    \caption{Predictions from model R-150 for the s-process elements abundance patterns vs. [Fe/H] for outer ($\mathrm{R_{GC}=12\ kpc}$), local ($\mathrm{R_{GC}=8\ kpc}$) and inner ($\mathrm{R_{GC}=6\ kpc}$) disc regions. The channels considered for the production of the s-process components are: massive stars with initial rotational velocities of $\mathrm{150\ km/s}$ and LIMS. The channels for the r-process components are: MR-SNe and MNS with a DTD. The curves are color coded by the age of the stars created by the chemical evolution code.}
    \label{fig: s_elements}
\end{figure*}

\subsubsection{s-process elements}

The s-process elements studied in this work are: Ba, La, Zr, Y and Ce. In Figure \ref{fig: Y and Ba obs only} we report the observational data for both the field stars and the OC sets at each galactocentric radius for [Y/Fe] and [Ba/Fe] vs. [Fe/H], taken as representative of the abundance pattern of s-process elements belonging to the first and second peak, respectively. A characteristic s-process elements `banana' shape is clearly seen at high metallicity, more pronounced in the [Ba/Fe] than in the [Y/Fe]. This shape is assumed to reflect the pollution from LIMS during the AGB phase, which enrich the ISM with s-process material at later times creating the peak at [Fe/H] $\mathrm{\sim -0.04\ dex}$. The decrease for higher metallicity values is then due to Fe production from SNeIa. This pattern is visible both in the field and in OCs. In particular, the OCs belonging to the outer disk have lower [Fe/H] and are characterized by an increasing pattern, the ones belonging to the solar region have nearly solar metallicity and display the peak followed by a slight decrease and finally the inner disk OCs, which have the highest [Fe/H], are characterized by a decreasing pattern, even if some of them overlaps with the trend defined by the solar-vicinity ones. 

According to \citet{prantzos2020} the fraction of the s-process elements studied in this work which is produced by the s-process is 78\%, 82\%, 89\%, 80\% and 85\% for Y, Zr, Ba, La and Ce, respectively, with the remaining fractions due mainly to the r-process (and in negligible or null amount to the p-process). Although the r-process fractions of those elements is not the predominant one, we will show that from a chemical evolution point of view it is necessary to include it as well, in order to reproduce the observed abundance trends. Therefore, as a first step, we focus on the results of our model obtained when no r-process nucleosynthesis is taken into account by showing what happens when only rotating massive stars and LIMS contribute to the s-process elements production. Results of our model together with the observed abundance patterns for the s-process elements Y and Ba are reported in Figure \ref{fig: s-elements solar RMS+reduced LIMS}. Outer, local and inner data are compared with our predicted evolution for $\mathrm{R_{GC}=12\ kpc}$, $\mathrm{R_{GC}=8\ kpc}$ and $\mathrm{R_{GC}=6\ kpc}$, respectively. The three different curves refer to the three different initial rotational velocities for massive stars: 0, 150 and 300 $\mathrm{km/s}$ (see Table \ref{tab: nucleosynthesis prescriptions}). The effect of rotation is clearly that of increasing the production of the s-process elements, especially for the elements belonging to the first s-process peak (Y in the Figure), as expected. For the elements which belong to the second s-process peak (Ba in the Figure), rotation must be increased to 300 $\mathrm{km/s}$ in order to see a significant enhancement in the stellar production. The second s-process production channel is represented by LIMS. With respect to previous chemical evolution studies, which adopted only yields from LIMS in the range $\mathrm{1.5-3.0\ M_\odot}$, here we extend the mass range to $\mathrm{1.5-8.0\ M_\odot}$. We adopt yields from \citet{Cristallo2009, Cristallo2011, Cristallo2015} up to $\mathrm{6.0\ M_\odot}$ and yields obtained by extrapolation in the range $\mathrm{6.0-8.0\ M_\odot}$. Moreover, since those yields tend to overproduce the solar abundances of the s-process elements, we reduced them by a factor of 2, as suggested by \citet{rizzuti2019} (see also \citealp{Rizzuti2021}). 

In Figure \ref{fig: s_elements} we report the results of model R-150 for the evolution of all the s-process elements studied in this work with both the s- and the r-process astrophysical sites activated. We are showing only the model with massive stars with initial rotational velocities of $\mathrm{150\ km/s}$ since, once the contribution from MR-SNe is considered, the differences in the predicted abundance patterns between model assuming $\mathrm{v_{rot}=0}$ and $\mathrm{150\ km/s}$ are negligible and, as previously shown, models with $\mathrm{v_{rot}=300\ km/s}$ overestimate the observed abundance trends. The predicted curves are color coded by the ages of the synthetic stars. We notice that the model is able to reproduce the main observed trends in the data, especially in the OC sample. The rise in the outer disc data as well as the peak followed by the decrease in the local and inner-disc data are reproduced by our model at $\mathrm{R_{GC}=12,\ 8}$ and $\mathrm{6\ kpc}$. The only exceptions are represented by the [Y/Fe] and the [Zr/Fe] vs. [Fe/H] trends in the outer region for which the model does not produce the expected increase, but rather a decrease. This is probably due to the combination of too high MR-SNe yields and too low LIMS ones for those two elements. Because of the MR-SNe yields, the model predicts high [Y/Fe] and [Zr/Fe] value at relatively low metallicities and it is not able to produce an increasing trend at higher ones because LIMS are not producing enough Zr and Y abundances. However, it must be reminded that LIMS are not supposed to be among the main producers of Y and Zr, since those two elements belong to the first s-process peak. [La/Fe] vs. [Fe/H] is slightly overproduced by the model in the local and inner disc regions. The [Ce/Fe] vs. [Fe/H] has recently been already studied by \citet{contursi2023} in the MW halo and disc component through a high quality samples of GSP-Spec Ce abundances. They find a rather flat trend at a mean level of [Ce/Fe]$\mathrm{\sim 0.2\ dex}$ for $\mathrm{-0.7<[M/H]<0.3\ dex}$ which are able to reproduce by means of the three-infall chemical evolution model by \citet{spitoni2023}. On the other hand, our OC sample clearly show the characteristic \textit{banana} shape of s-process elements, rather than a flat trend, which is clearly well reproduced by our two-infall model. The three curves reach metallicities of [Fe/H] $\mathrm{\sim -0.16,\ 0.16}$ and $\mathrm{0.33\ dex}$ in the outer, local and inner-disc, respectively, which slightly underestimate the ones observed in the OC samples for the outer and local disc, similarly to what happen for Eu (see previous Section).
%We remind that for the r-process component we use i) 20\% of massive stars with progenitors mass in the $\mathrm{10-25\ M_\odot}$ mass range which explode as MRD-SNe with yields from \citet{Nishimura2017} (model L0.75) and ii) MNS with a DTD from \cite{Simonetti2019} (with $\mathrm{\beta=-0.9}$) and with yields obtained by scaling the one of Sr measured by \citet{Watson2019}. For the s-process component we use i) LIMS with $\mathrm{1.5-8.0\ M_\odot}$ with yields from \citet{Cristallo2009, Cristallo2011, Cristallo2015} (non rotating case and manually reduced by a factor 2) and ii) rotating massive stars ($\mathrm{M>8\ M_\odot}$) with yields of \citet{Limongi2018} with initial rotational velocity of $\mathrm{150\ km/s}$. For the (see Section \ref{sec: nucleosynthesis prescriptions}).  

\begin{figure*}
    \centering
    \includegraphics[width=1\textwidth]{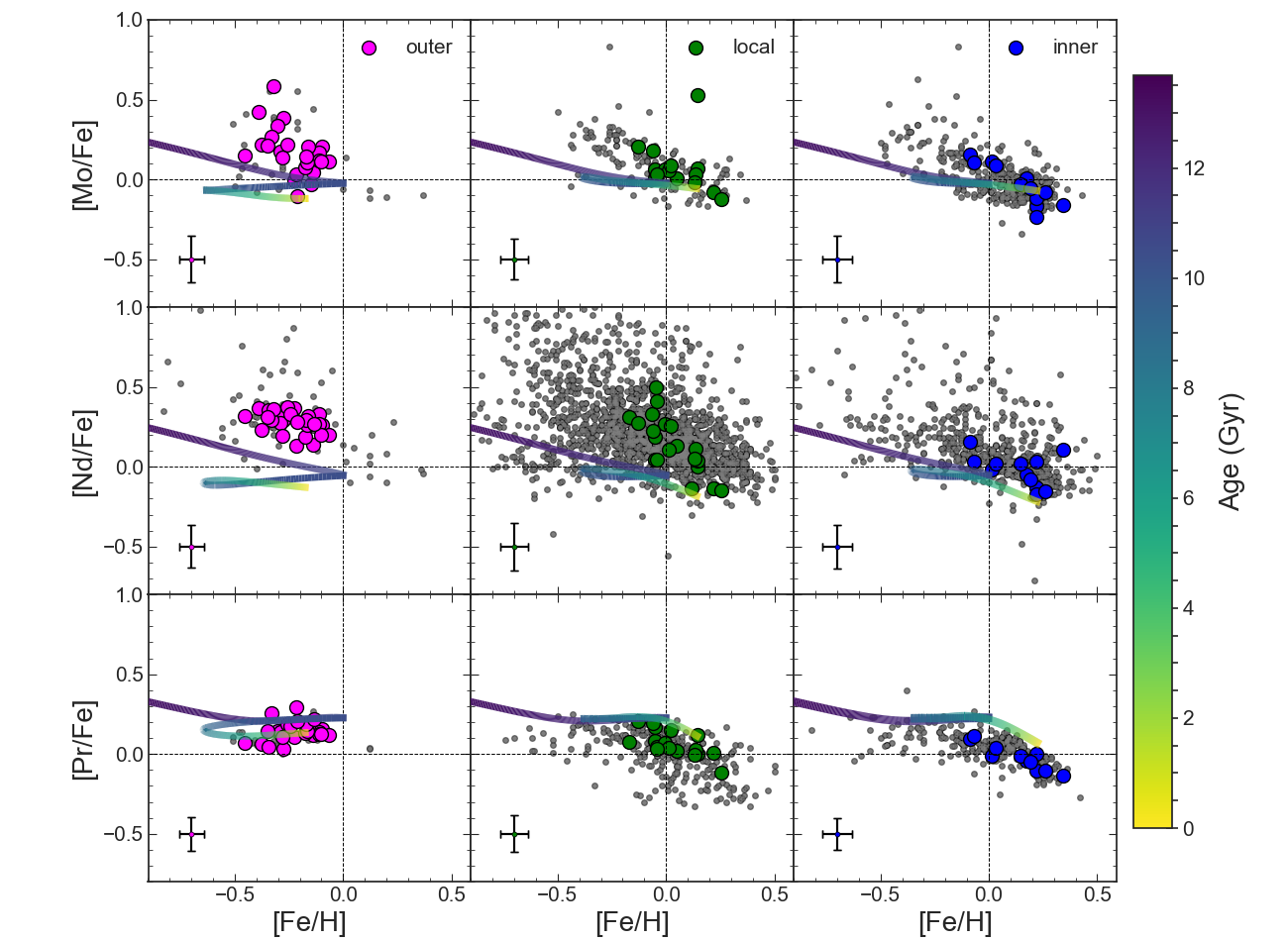}
    \caption{Predictions from model R-150 for the mixed-process elements abundance patterns vs. [Fe/H] for outer ($\mathrm{R_{GC}=12\ kpc}$), local ($\mathrm{R_{GC}=8\ kpc}$) and inner ($\mathrm{R_{GC}=6\ kpc}$) disc regions. The curves are color coded by the age of stars created by the chemical evolution code.}
    \label{fig: r_elements}
\end{figure*}

\subsubsection{Mixed-process elements}
\label{sec: mixed elements}

In Figure \ref{fig: r_elements}, we report the observed abundance patterns for both field stars and OCs in our \textit{Gaia}-ESO DR6 samples (see Section \ref{sec: observational data}) together with the predictions from our model for Mo, Nd and Pr. We refer to those elements as mixed process elements. In fact, even if for all the elements studied in this work both the contributions from the s- and the r-process have been considered, Mo, Nd and Pr are found to owe a large fraction of their Galactic abundances to the r-process. As also discussed by \citet{vanderswaelmen2023}, different studies (e.g. \citealp{sneden2008, bisterzo2014, prantzos2020}) agree in assigning $\sim40\%$ of the r-process component to Nd and $\sim 50\%$ to Pr. On the other hand, the contributions of the different processes to the abundance of Mo in the Sun differ from one author to another. According to \citet{Hansen2014}, who presented a study of both Mo and Ru abundances in the MW covering both dwarfs and giants from [Fe/H] $\mathrm{\sim -0.7}$ down to $\mathrm{\sim -3.2\ dex}$, Mo can be considered as a highly mixed element, with contributions from the main and weak s-processes as well as from the p-process and, in a smaller fraction, from the main r-process. In agreement with that study, more recently \citet{prantzos2020} proposed for Mo a contribution of $50\%$ from s-process, $27\%$ from r-process and $23\%$ from p-process. \citet{vanderswaelmen2023}, which adopted our same data set, examined also the origin of these elements from an observational point of view, comparing their abundance with that of Eu. From their study, it appears that Nd should be characterized by a significant s-process contribution, whereas for Pr they expect a lower contribution from the s-process. Results of our model are not in agreement with these hypotheses for Nd and Pr. In fact, as shown in Figure \ref{fig: r_elements},  our model predicts a too low [Nd/Fe] vs. [Fe/H] abundance pattern with respect to the observed one. This discrepancy between predictions and observations may be attributed to LIMS which produce a too low amount of Nd with respect to what is expected in \citet{vanderswaelmen2023}. On the other hand, our model fits properly the [Pr/Fe] vs. [Fe/H] trend, only slightly overproducing the observed pattern in the local and inner parts. This may be due both to a too strong production of Pr by massive stars and by LIMS, contrary to what happens for Nd. It is worth noting that the observed abundance pattern of [Pr/Fe] vs. [Fe/H] resembles more that of an s-process elements than the one of the [Nd/Fe] vs. [Fe/H], at least when considering the \textit{Gaia}-ESO OC data. In fact, in the case of Pr we distinguish a rise in the abundances in the outer region, followed by a peak and then a decrease in the local and inner regions. On the other hand, this typical 'banana' shape is not recognisable in the observed abundance pattern of Nd.

The behaviour of Mo is much more uncertain. As discussed in \citet{vanderswaelmen2023}, the elusive nature of this chemical element together with the difficulty in measuring its abundance caused chemical evolution studies to reach discordant conclusions about its cosmic origin. In \citet{Mishenina2019}, the [Mo/Fe] vs. [Fe/H] abundance pattern has been studied in a wider range of metallicities with respect to our samples and with chemical evolution models from \citet{Travaglio2004}, \citet{Prantzos2018} and with the open-source galactic chemical evolution code OMEGA+ (\citealp{cotè2018}). Their main conclusion is that canonical stellar sources of heavy elements are not producing a sufficient amount of Mo to reproduce observations. They showed that, despite the fact that the r-process contributes to a small fraction of the solar Mo, it is of significant importance especially at low metallicities, where the s-process contribution from AGB stars is negligible. In fact, in \citet{Mishenina2019} the model that better agrees with the data is the one in which the r-process component of Mo is produced on quick timescales (the r-process production site considered is associated with MNS with a short and constant delay time for merging). Our results for the [Mo/Fe] vs. [Fe/H] abundance pattern are in agreement with \citet{Mishenina2019} conclusions. In fact, even with a quick r-process source activated (in our case represented by MR-SNe) the predicted trends appear to slightly underestimate the observed ones at each Galactocentric radius. However, this is most probably due to the lack in our model of an additional contribution from neutrino-driven SNe which may be important producers of Mo at all metallicities (see e.g. \citealp{Bliss2018,Bliss2020}).

\subsubsection{Comparison with previous studies}

Similar prescription to those adopted in this work have been already included in chemical evolution studies by \citet{Prantzos2018} (from now on P18) and \citet{rizzuti2019} (from now on R19) in order to study the contribution from rotating massive stars to the enrichment of different chemical species in the MW. The main differences between the work of R19 and this one are: i) the set of \citet{Limongi2018} for rotating massive stars used by R19 is the Set F whereas we adopt \citet{Limongi2018}'s recommended Set R (see also \citealp{romano2019}); ii) in R19 LIMS in the $\mathrm{1.3-3.0\ M_\odot}$ mass range are assumed to produce s-process elements, whereas we extend the mass range to $\mathrm{1.3-8.0\ M_\odot}$; iii) for the r-process nucleosynthesis we use both MR-SNe and MNS, with these latter characterized by having a DTD, while R19 used either one source or the other (with the coalescence time-scale for MNS constant and equal to $\mathrm{1\ Myr}$); iv) the iron yields from CC-SNe adopted in R19 are those from \citet{kobayashi2006}, while for consistency here we adopt the ones from \citet{Limongi2018}. On the other hand, the main differences between the work of P18 and our are: i) in P18 r-process elements are assumed to be produced in CC-SNe and their yields are scaled to the yield of oxygen according to the solar system r-process contribution as determined by \citet{sneden2008}. The yields so obtained, are functions of the mass and of the metallicity of the star; ii) the yields of rotating massive stars are weighted with a metallicity dependent function empirically determined.

The main difference between our results and those of R19 is that in their model it appears that the contribution from MR-SNe is not the dominant one at really low metallicities. In fact, it is still possible to appreciate differences between their model in which massive stars have an initial rotational velocity of $\mathrm{150\ km/s}$ and the one in which massive stars do not rotate at all, even when the MR-SNe channel is active (their models LC000+MRD and LC150+MRD). These differences are visible only at really low metallicities ($\mathrm{<-4\ dex}$). For higher [Fe/H] the two models are very similar, exactly as it happens in our case. So that, for the metallicities we are interested in this work, we do not expect significant differences. The dissimilarities in the prescriptions adopted for MR-SNe nucleosynthesis between our work and that of R19 may be responsible for the slightly different results between the two studies.

The discrepancies between our predictions and those of P18 are not so strong, even if their prescriptions for the production of heavy elements from massive stars is substantially different from ours. In fact, they assume that all CC-SNe can produce heavy elements and scale their yields to the one of oxygen, whereas in our case only a small fraction of massive stars can produce r-/s- process material with the nucleosynthesis of \citet{Nishimura2017}. The major issue which results from our approach is that the adopted yields are not a function of the metallicity and, in particular, of the mass of the progenitor. That means that all the assumed 20\% of stars with progenitor mass of $\mathrm{10-25\ M_\odot}$ are producing the same amount of r- process material independently from the metallicity, which is of course an oversimplification. P18 assumed that the yield of each heavy element considered scales with another element produced exclusively by massive stars which reproduce the solar abundance, with this latter being already a function of mass and metallicity. This method still has some uncertainties but it can be reliable for the purpose of P18, which is the study of the effect of rotating massive stars yields. In our case, the primary goal is that of studying the origin of neutron capture elements by adopting the state-of-the-art in the nucleosynthesis prescriptions and, as a result, to reveal the main uncertainties in both the chemical evolution models and the nucleosynthesis itself.

\begin{figure*}
    \begin{center}   \subfloat{\includegraphics[width=1\columnwidth]{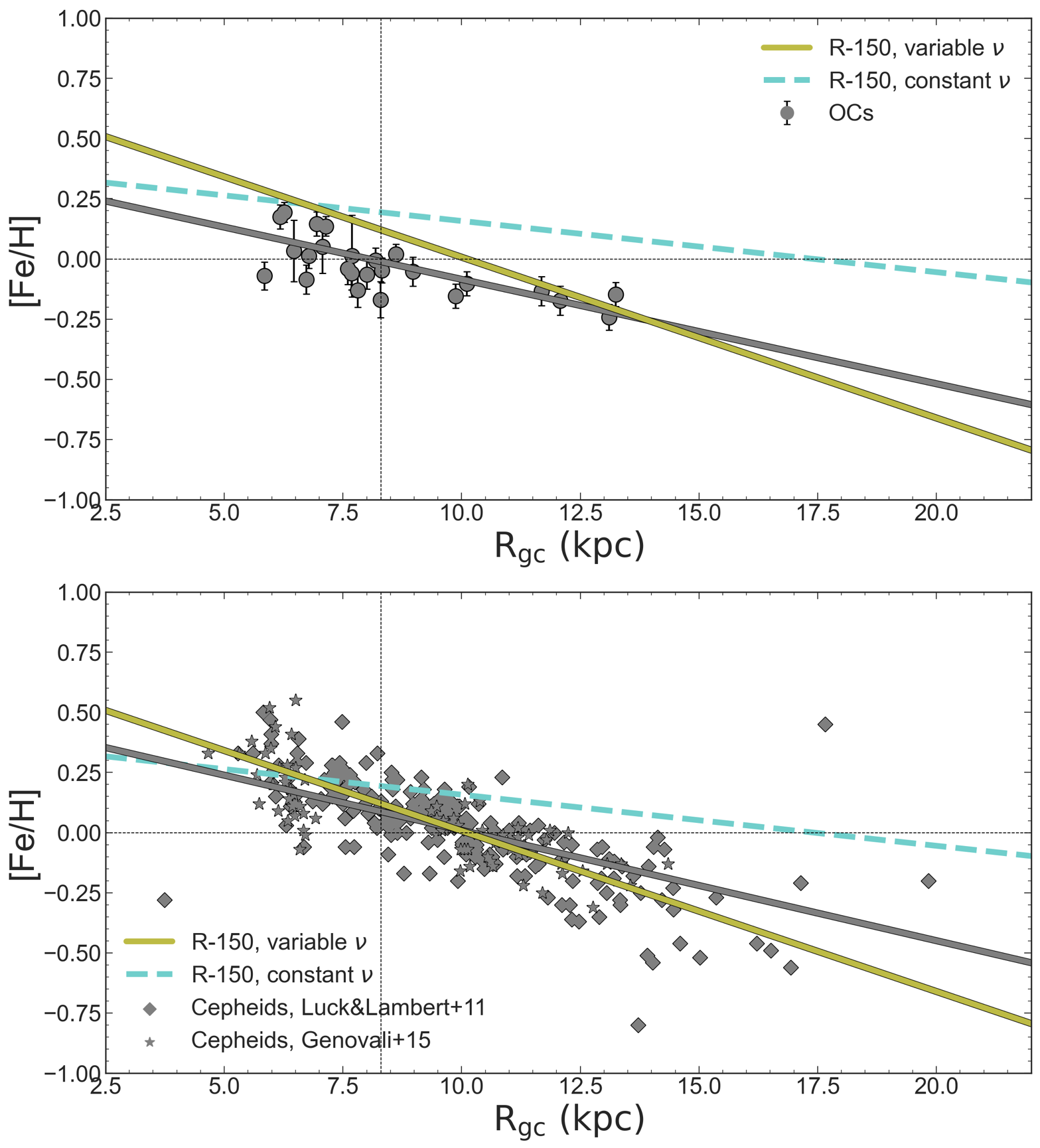}\label{fig:a}}     \hspace{0.4cm}\subfloat{\includegraphics[width=1\columnwidth]{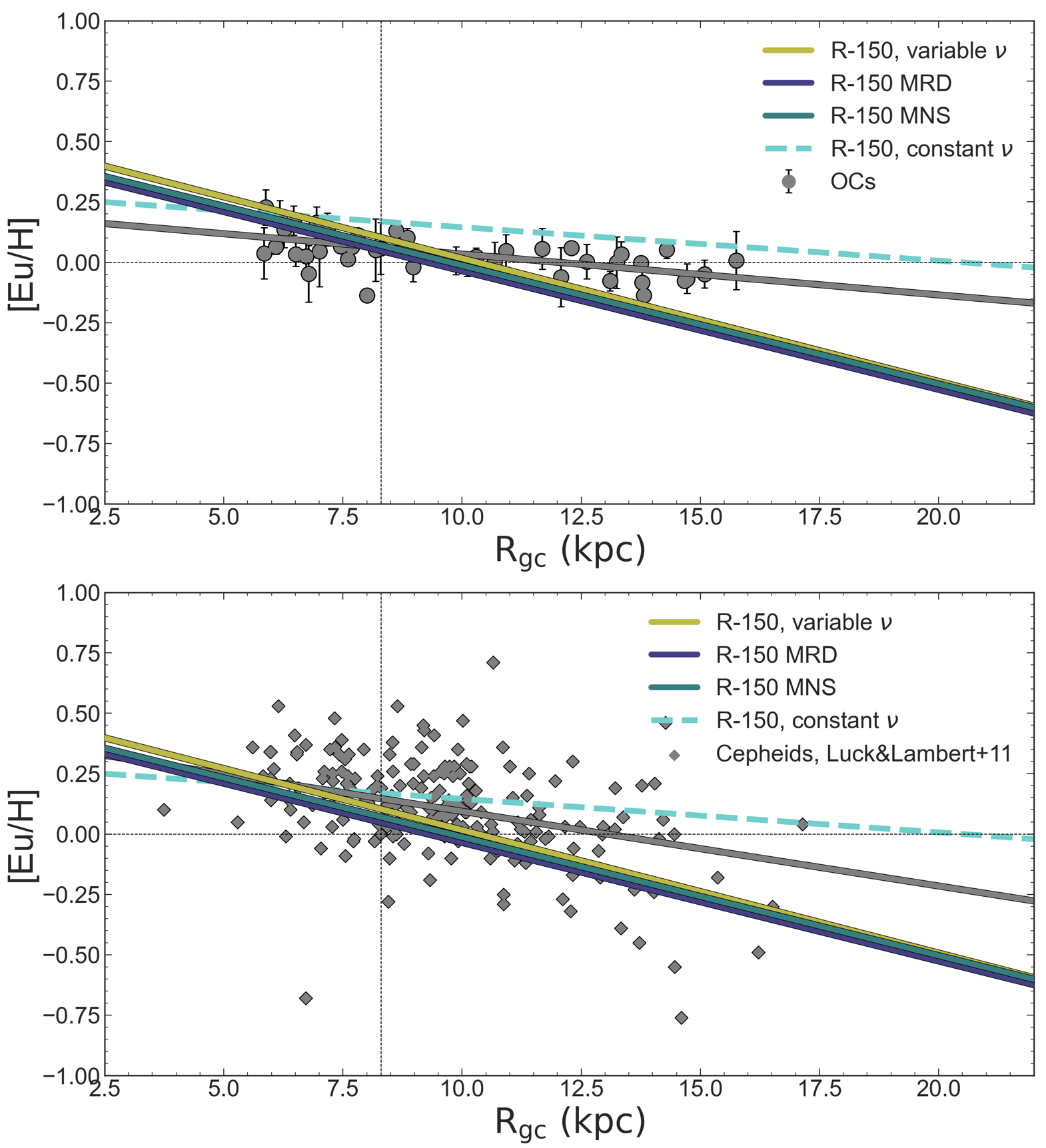}\label{fig:a}}
    \caption{Prediction of the present day slope of the [Fe/H] and [Eu/H] gradients from our models with variable (olive line) and constant (dashed turquoise line) SF efficiency compared to the the one of the restricted ($\mathrm{Age\ \le3\ Gyr}$) OC sample (upper panel) and the Cepheid sample (lower panel) with data from \citet{Luck2011} (grey diamonds) and \citet{Genovali2015} (grey stars). For the [Eu/H] we show also models with Eu produced only by MR-SNe (purple line) and Eu produced only by MNS with a constant and short delay time for merging (teal line). The grey lines represent the linear fit of the observational data.}%
    \label{fig: gradients linear fit Fe/Eu}%
    \end{center}
\end{figure*}

%\begin{figure}
%    \begin{center}
%    \includegraphics[width=1\columnwidth]{Images/gradients/EuH_merged.png}
%    \caption{Prediction of the present day slopes of the [Eu/H] gradient from three different models (Eu produced by both MNS and MR-SNe with either a variable or a constant SF efficiency, olive and turquoise lines; Eu produced only by MR-SNe, purple line; Eu produced only by MNS with a constant and short delay time for merging, teal line) compared to the the one of the restricted ($\mathrm{Age\ \le3\ Gyr}$) OCs sample (upper panel) and the Cepheids sample (lower panel) with data from \citet{Luck2011} (grey diamonds).}%
%    \label{fig: gradients linear fit Eu}%
%    \end{center}
%\end{figure}

\subsection{Gradients}
\label{sec: gradients}

Here, we present the abundance gradients of the studied elements. We first show our predictions for the present time abundance gradients along the disc and then discuss their time evolution.

\subsubsection{Present day radial abundance gradients}

In Figure \ref{fig: gradients linear fit Fe/Eu} we compare the theoretical present-day gradients of [Fe/H] and [Eu/H] to the observational data. In order to compare present day results of our model with the observations, we restricted our OC sample to clusters with $\mathrm{Age\ \le3\ Gyr}$. As discussed in \citet{Magrini2023} (from now on M23; see also references therein) there is a general agreement about the existence of a steeper [Fe/H] gradient in the inner disc and an extended plateau in the outer region, with a cutoff point at $\mathrm{R_{GC}\sim11.2\ kpc}$. This change of slope is evident in the OC sample used in this work and it is still visible when we restrict our sample to OCs younger than 3 Gyr. Considering the entire OC sample and a weighted single slope fit, M23 find a slope of the [Fe/H] gradient of $\mathrm{-0.054\pm0.004\ dex\ kpc^{-1}}$. Considering the two radial region they obtain a steeper inner gradient ($\mathrm{-0.081\pm0.008\ dex\ kpc^{-1}}$) and a much flatter outer plateau ($\mathrm{-0.044\pm0.014\ dex\ kpc^{-1}}$). The slopes of the restricted OC sample are in good agreement with the ones of the whole sample, as reported in Table \ref{tab: slopes of the [Fe-Eu/H] gradients}. The slope of the [Fe/H] gradient predicted by our model is  equal to $\mathrm{-0.067\pm0.003\ dex\ kpc^{-1}}$. As shown in Figure \ref{fig: gradients linear fit Fe/Eu}, this is slightly steeper both with respect to the gradients of the restricted OC sample (upper panel) and to the one of the Cepheid sample including the data of \citet{Luck2011} and \citet{Genovali2015} (lower panel). However, our result is in good agreement with other recent literature slopes of the [Fe/H] gradient from OC samples, (\citealp{Carrera2019,Donor2020,Zhang2021,Spina2021,spina2022}, see Table 1 in M23). 

In the right panels of Figure \ref{fig: gradients linear fit Fe/Eu} we compare the slopes predicted by our model under different assumptions (see Table \ref{tab: nucleosynthesis prescriptions}) for the [Eu/H] gradient with those observed in the restricted OC sample and in the Cepheid one from \citet{Luck2011}. The OC restricted sample shows a flat [Eu/H] gradient, with a global slope equal to $\mathrm{-0.017\pm0.004\ dex\ kpc^{-1}}$ and an inner slope slightly steeper than the outer one (see Table \ref{tab: slopes of the [Fe-Eu/H] gradients}). In general, elements which are produced on longer timescales are characterized by steeper gradients than elements produced on fast timescales. For example, $\alpha$-elements have flatter slopes than Fe-peak elements (even if there may be differences also between elements of the same group). Therefore, the shape of the [Eu/H] gradient points towards a short timescale of production. We remind that in model R-150 (variable $\nu$) the r-process material comes from both a quick source (MR-SNe) and a delayed one (MNS with a DTD). In Figure \ref{fig: gradients linear fit Fe/Eu} this model is represented by the olive line. It predicts a global slope equal to $\mathrm{-0.051\pm0.003\ dex\ kpc^{-1}}$, which is too steep with respect to the one observed from the OC sample. The agreement with the data does not improve even if we assume that Eu is produced only on short timescales (purple and teal lines in the Figure, corresponding to Eu production from solely MR-SNe and MNS with a constant and short ($\mathrm{10\ Myr}$) delay time for merging, respectively). Moreover, these two models predict the same slope of the [Eu/H] gradient, equal to $\mathrm{-0.049\pm0.003\ dex\ kpc^{-1}}$. Our results do not improve noticeably even if the contribution from the delayed source is suppressed, because MNS with DTD are not the main source of r-process material in model R-150. In fact, even if their Eu yield is higher with respect to that of MR-SNe, their rate is low (see Figure \ref{fig:rate}). We stress once again that MNS are the only source of r-process material confirmed by observations and therefore they must be included in the computation. However, when compared with the MR-SNe, they are not the dominant source. On the other hand, by including them the agreement between our predicted [Eu/H] inner slope and the observed one improves, especially for models with no delayed source (see Table \ref{tab: slopes of the [Fe-Eu/H] gradients}).

To reproduce flatter present-day abundances gradients we tested also a model with constant SF efficiency ($\mathrm{\nu=1\ Gyr^{-1}}$). Results for the [Fe/H] and for the [Eu/H] are reported in Figure \ref{fig: gradients linear fit Fe/Eu} as well as in Table \ref{tab: slopes of the [Fe-Eu/H] gradients}. By assuming a constant SF efficiency we obtain a satisfactory agreement with the OCs [Eu/H] gradient. On the other hand, we lose the agreement with the observed [Fe/H] gradient, in particular in the outer regions where the predicted SF turns out to be too intense. As already pointed out by \citet{Grisoni2018}, the inside-out scenario, although is a key ingredient for the formation of the Galactic discs, is not enough to explain the abundance pattern at different Galactocentric distances and the abundance gradients by itself. Models with only an inside-out scenario usually predicts too flat present day gradients, as is the case of our model with constant SF efficiency. In order to steepen the gradients further assumptions are needed. In particular, one need to consider either a variable SF efficiency, or radial gas flows or a combination of both (\citealp{Palla2020}). Models with decreasing SF efficiency with increasing Galactic radius produce a steeper gradient, since they boost the chemical enrichment in the inner regions relative to the outer ones. Radial migration of stars, which is not taken into account in any of our models, on the contrary, should have the effect of flattening the gradient on long enough timescales (\citealp{Minchev2018,Quillen2018}). Whether clusters are affected by migration as much as field stars is not completely understood yet. If also clusters with $\mathrm{Age<3\ Gyr}$ are affected by migration, the discrepancy between our models and the data (especially in the outer region) may be partially due to the moving outward of "old" clusters formed in the inner disc (see M23 and references therein). In favor of this hypothesis, the slopes of the [Eu/H] gradient predicted by our models with variable SF efficiency are in better agreement with the one computed from the younger Cepheid sample of \citet{Luck2011}, shown in the lower panels of Figure \ref{fig: gradients linear fit Fe/Eu} (see Table \ref{tab: slopes of the [Fe-Eu/H] gradients}).

\begin{table*}
\centering
\begin{tabular}{lccccccc}
\hline
   &                    &     [Fe/H]           &                     &  &  &[Eu/H] &  \\
\hline
   & $\mathrm{m_{tot}}$ & $\mathrm{m_{inner}}$ & $\mathrm{m_{outer}}$ & & $\mathrm{m_{tot}}$ & $\mathrm{m_{inner}}$ & $\mathrm{m_{outer}}$ \\
   & $\mathrm{(dex\ kpc^{-1})}$ & $\mathrm{(dex\ kpc^{-1})}$ & $\mathrm{(dex\ kpc^{-1})}$ & & $\mathrm{(dex\ kpc^{-1})}$ & $\mathrm{(dex\ kpc^{-1})}$ & $\mathrm{(dex\ kpc^{-1})}$ \\
\hline
  OCs & $\mathrm{-0.049\pm0.005}$ &  $\mathrm{-0.081\pm0.013}$ & $\mathrm{-0.045\pm0.017}$ & & $\mathrm{-0.017\pm0.003}$ & $\mathrm{-0.024\pm0.009}$  &  $\mathrm{-0.015\pm0.014}$\\
  Cepheids & $\mathrm{-0.046\pm0.003}$ &  - & - & & $\mathrm{-0.031\pm0.004}$ &  - & - \\
\hline
  Model R-150 var $\nu$ & $\mathrm{-0.067\pm0.002}$ & $\mathrm{-0.064\pm0.008}$ &  $\mathrm{-0.063\pm0.007}$& & $\mathrm{-0.051\pm0.003}$ & $\mathrm{-0.038\pm0.004}$ & $\mathrm{-0.057\pm0.007}$ \\
  Model R-150 con $\nu$ & $\mathrm{-0.021\pm0.004}$ & $\mathrm{-0.044\pm0.008}$ & $\mathrm{-0.007\pm0.001}$ & & $\mathrm{-0.014\pm0.003}$ & $\mathrm{-0.029\pm0.006}$ & $\mathrm{-0.004\pm0.001}$ \\
  Model R-150 MRD & - & - & - & & $\mathrm{-0.049\pm0.003}$ & $\mathrm{-0.034\pm0.003}$ & $\mathrm{-0.057\pm0.007}$ \\
  Model R-150 MNS  & - & - & - & & $\mathrm{-0.049\pm0.003}$ & $\mathrm{-0.034\pm0.003}$ & $\mathrm{-0.057\pm0.007}$ \\

\hline
\end{tabular}%
\caption{\label{tab: slopes of the [Fe-Eu/H] gradients} Slopes of the [Fe/H] and [Eu/H] gradients of the reduced ($\mathrm{Age\le3\ Gyr}$) OC sample and as predicted by our model for the all, inner ($\mathrm{R_{GC}<11.2\ kpc}$) and outer ($\mathrm{R_{GC}>11.2\ kpc}$) radial region. For comparison we show also the one obtained from the Cepheid sample of \citet{Luck2011} and \citet{Genovali2015}. In the case of Eu we show predictions also of models in which Eu is produced either by MR-SNe or by MNS with a constant and short delay time for merging.}
\end{table*}

\begin{figure*}
    \begin{center}
    \subfloat{\includegraphics[width=1\textwidth]{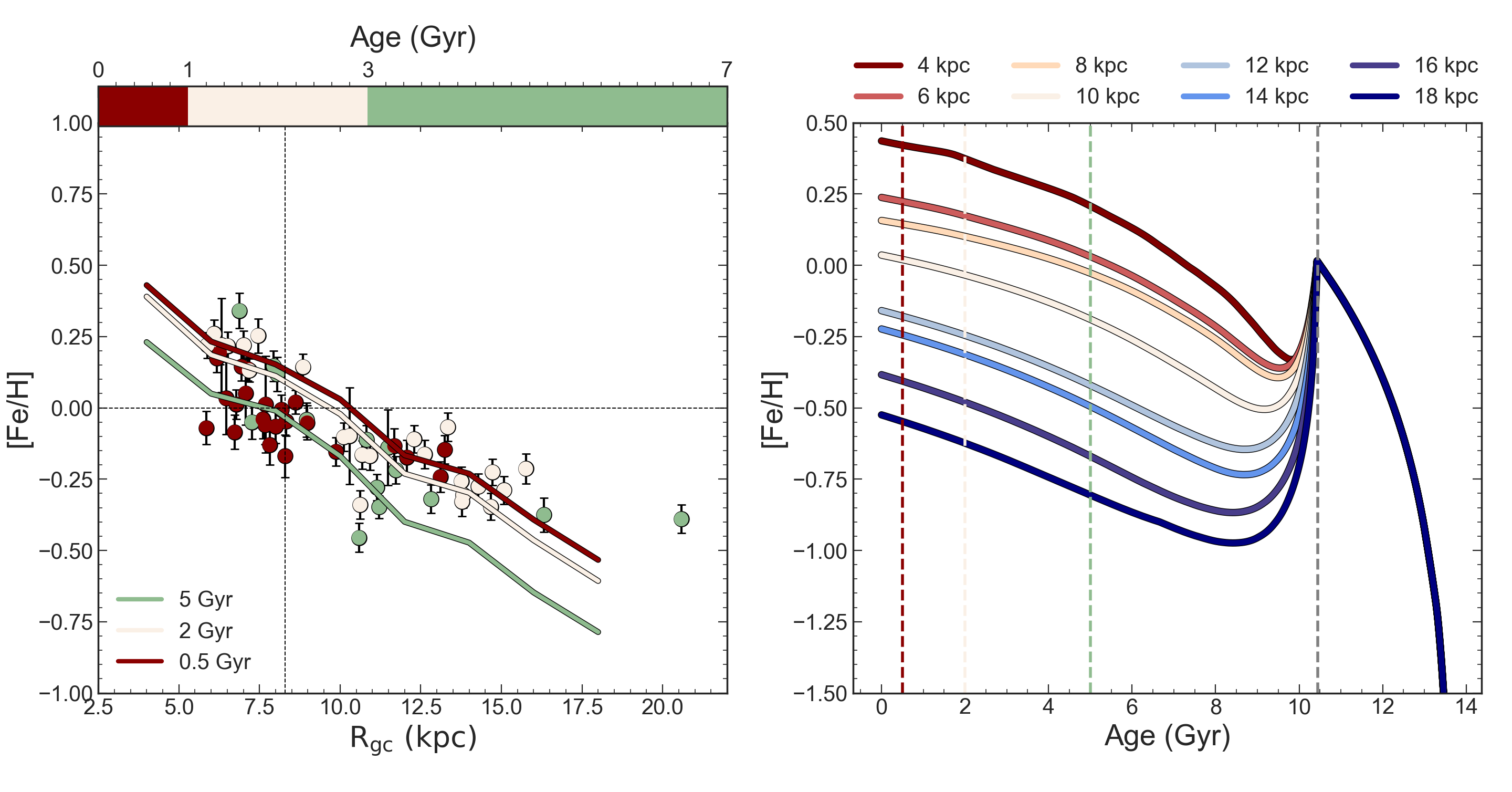}\label{fig:a}}
    \vspace{-0.6cm}
    \subfloat{\includegraphics[width=1\textwidth]{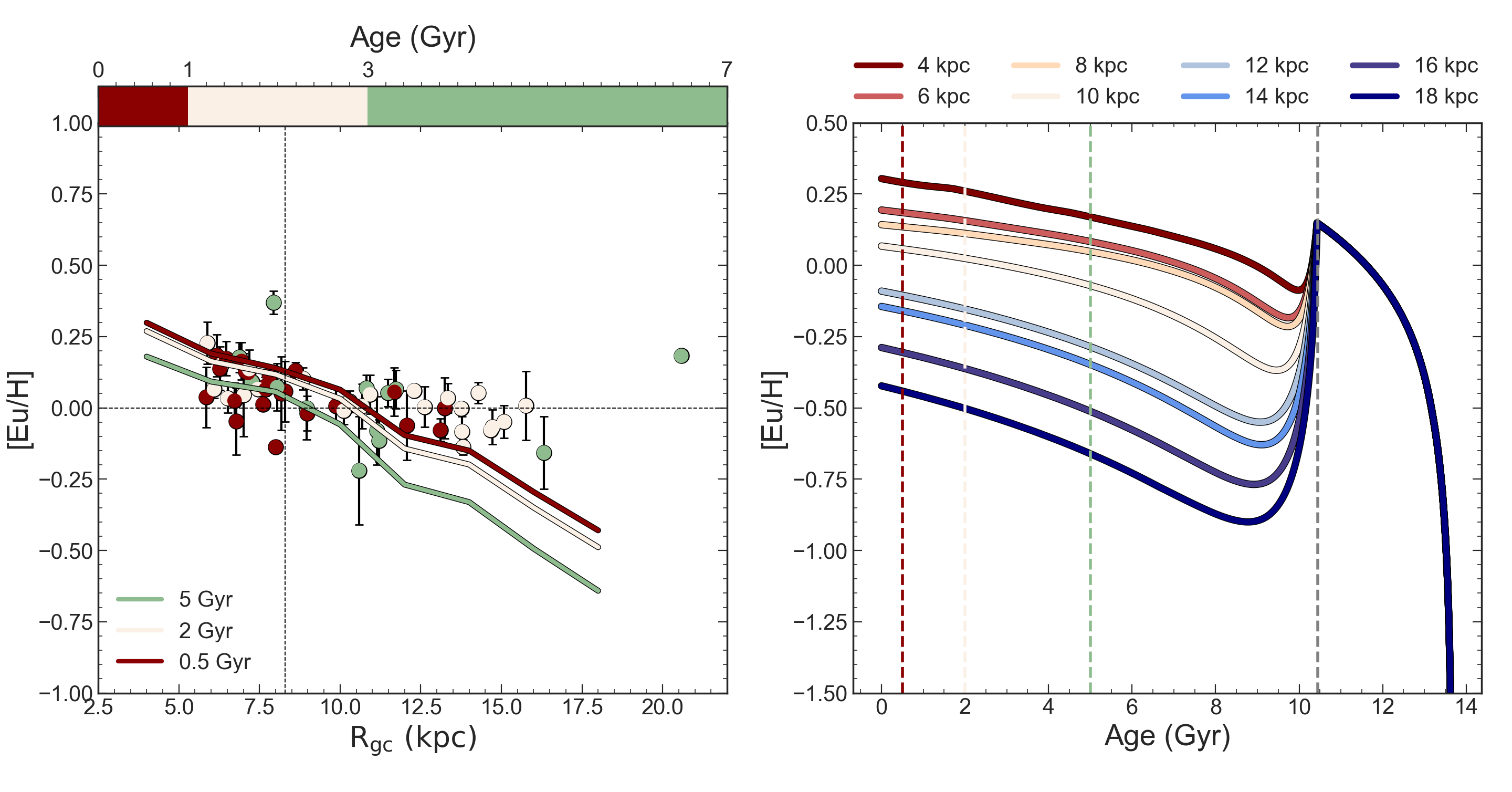}\label{fig:b}}%
    \caption{Left panels: Time evolution of the radial [Fe/H] and [Eu/H] gradients as predicted by model R-150. The OC sample is divided in three age bins: young ($\mathrm{Age<1\ Gyr}$), intermediate ($\mathrm{1<Age<3\ Gyr}$) and old ($\mathrm{Age>3\ Gyr}$). Solid lines are the results for the [Fe/H] gradient as predicted by our model at $\mathrm{0.5\ Gyr}$, $\mathrm{2\ Gyr}$ and $\mathrm{5\ Gyr}$. Right panel: time evolution of the [Fe/H] and [Eu/H] as predicted by model R-150 for different Galactocentric distances. Vertical dotted lines indicate the ages considered to compute the gradients.}
    \label{fig: FeH gradients}%
    \end{center}
\end{figure*}

\subsubsection{Time evolution of the radial abundance gradients}

In the upper left panel of Figure \ref{fig: FeH gradients}, we report the time evolution of the radial [Fe/H] gradients of the completed OC sample divided in three age bins together with results of our model at $0.5$, $2$ and $\mathrm{5\ Gyr}$. In the upper right panel of the same Figure we report the predictions about [Fe/H] evolution as a function of Age for different radii. The drop at $\mathrm{Age\sim10.44\ Gyr}$, in correspondence of the vertical grey dotted line, is the effect of the dilution event which happens when the second infall forming the thin disc takes place. Since we are comparing our results with OCs in the thin disk, we are interested only in the evolution from $\mathrm{Age\sim10.55\ Gyr}$ until the present day.

As observed by M23, the youngest clusters of the sample ($\mathrm{Age<1\ Gyr}$) have lower metallicity than the older ones in the inner disc ($\mathrm{R_{GC}<10\ kpc}$). As expected, the trend in the youngest clusters is not in agreement with our chemical evolution simulations which predict that the oldest population should be less enriched than the youngest one (on the other hand, an additional recent third infall episode produces a chemical impoverishment of the young population; see \citealp{spitoni2023}). The young clusters also show a flatter [Fe/H] gradient with a slope of $\mathrm{-0.038\pm0.004}$ (for $\mathrm{Age<1\ Gyr}$), $\mathrm{-0.063\pm0.006}$ (for $\mathrm{1\leq Age\leq3\ Gyr}$) and $\mathrm{-0.084\pm0.019}$ (for $\mathrm{Age>3\ Gyr}$) (see Table A.10 of M23). The slopes predicted by our model at $\mathrm{5}$, $\mathrm{2}$ and $\mathrm{0.5\ Gyr}$ reproduce this trend, but the difference between the three slopes is not that significant (see Table \ref{tab: gradients with ages}). Larger variations with time of the gradient slopes would be obtained by comparing our model results at older times. On the other hand, really small changes are expected in the latest Gyrs, as it appears clear from the upper right panel of Figure \ref{fig: FeH gradients}. According to M23, the observed trend in the youngest clusters is most likely due to a bias introduced by the standard spectroscopic analysis of low gravity giant stars. If the gradient of the youngest population is recomputed by removing giant stars with $\mathrm{log}g\mathrm{<2.5}$ the final gradient is very close to that of OCs with $\mathrm{1<Age<3\ Gyr}$ which suggests a limited time evolution of the gradient, in agreement with our models.

The lower panels of Figure \ref{fig: FeH gradients} are the same as the upper panels, but for Eu. Here, we show results of our model R-150. Unlike [Fe/H], in the case of Eu the OC sample gradient does not show different shapes with time. The youngest population shows abundances consistent with that of the intermediate and older clusters. Our model is in agreement with this trend, in fact it predicts very similar slopes for the three different lines corresponding to ages of $0.5$, $2$ and $\mathrm{5\ Gyr}$ (see Table \ref{tab: gradients with ages}). As already discussed previously, we predict a steeper present day slope with respect to the observed one. This is true also for the different ages shown in Figure \ref{fig: FeH gradients}. However, as for the present day gradients, also at different ages we obtain a much better agreement with the data of the inner ($\mathrm{R_{GC}<11.5\ kpc}$) disc rather than with those of the outer parts, where our model struggles to reproduced the observed plateau. This could be due to the too low SF efficiencies assumed for the outer part of the disc. However, it must be pointed out that we do not expect much higher SF in the outer disc and the SF efficiency has been fine tuned to reproduce the abundance patterns of the OCs with $\mathrm{R_{GC}>9\ kpc}$, as shown in the previous sections. Higher values of the SF efficiency would produce a too high metallicity and the agreement with both the abundances pattern and the [Fe/H] gradients would be lost. 

\begin{figure}
    \begin{center}
    \subfloat{\includegraphics[width=1\columnwidth]{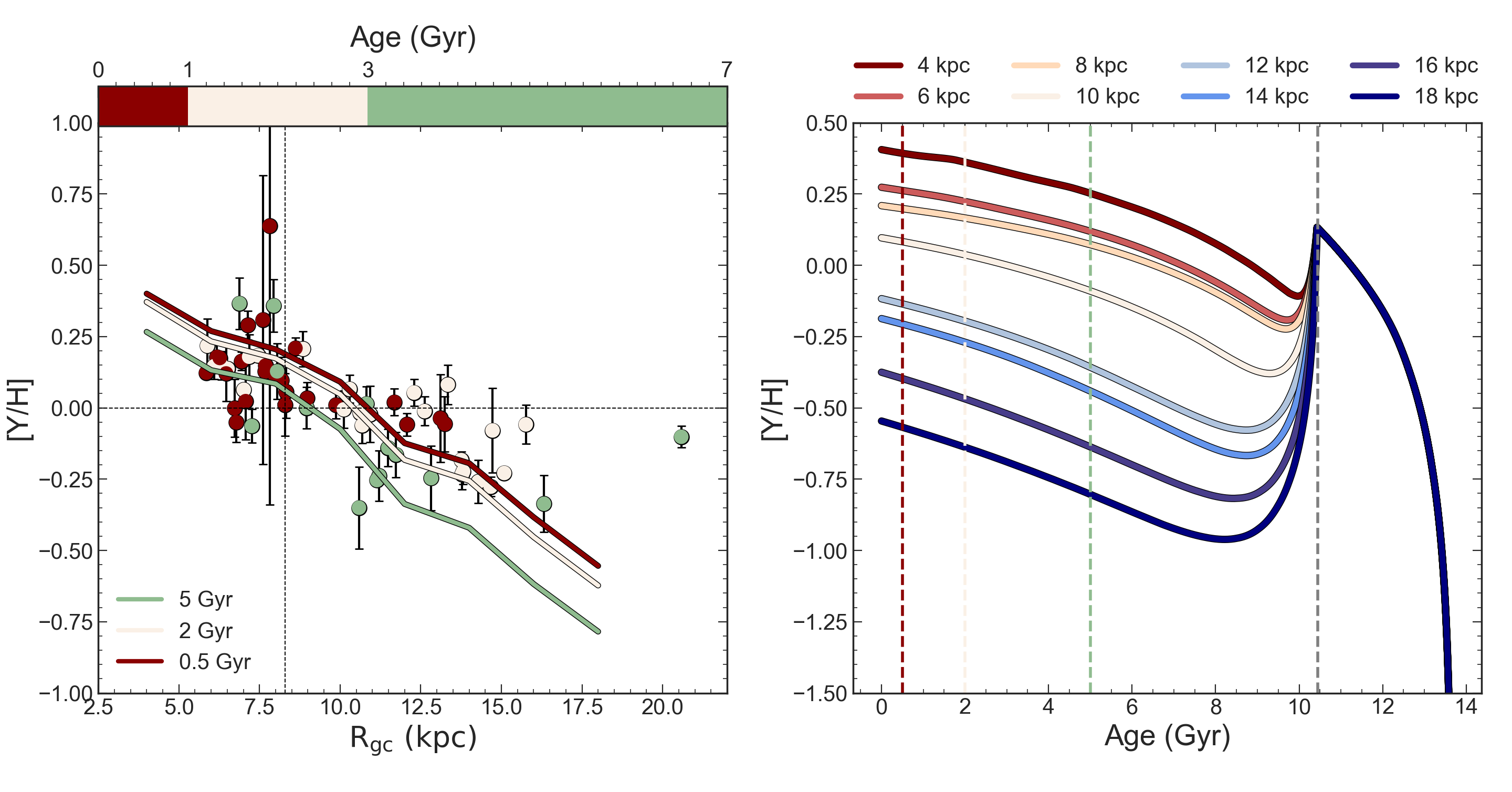}\label{fig:a}}
    \hfill
    \subfloat{\includegraphics[width=1\columnwidth]{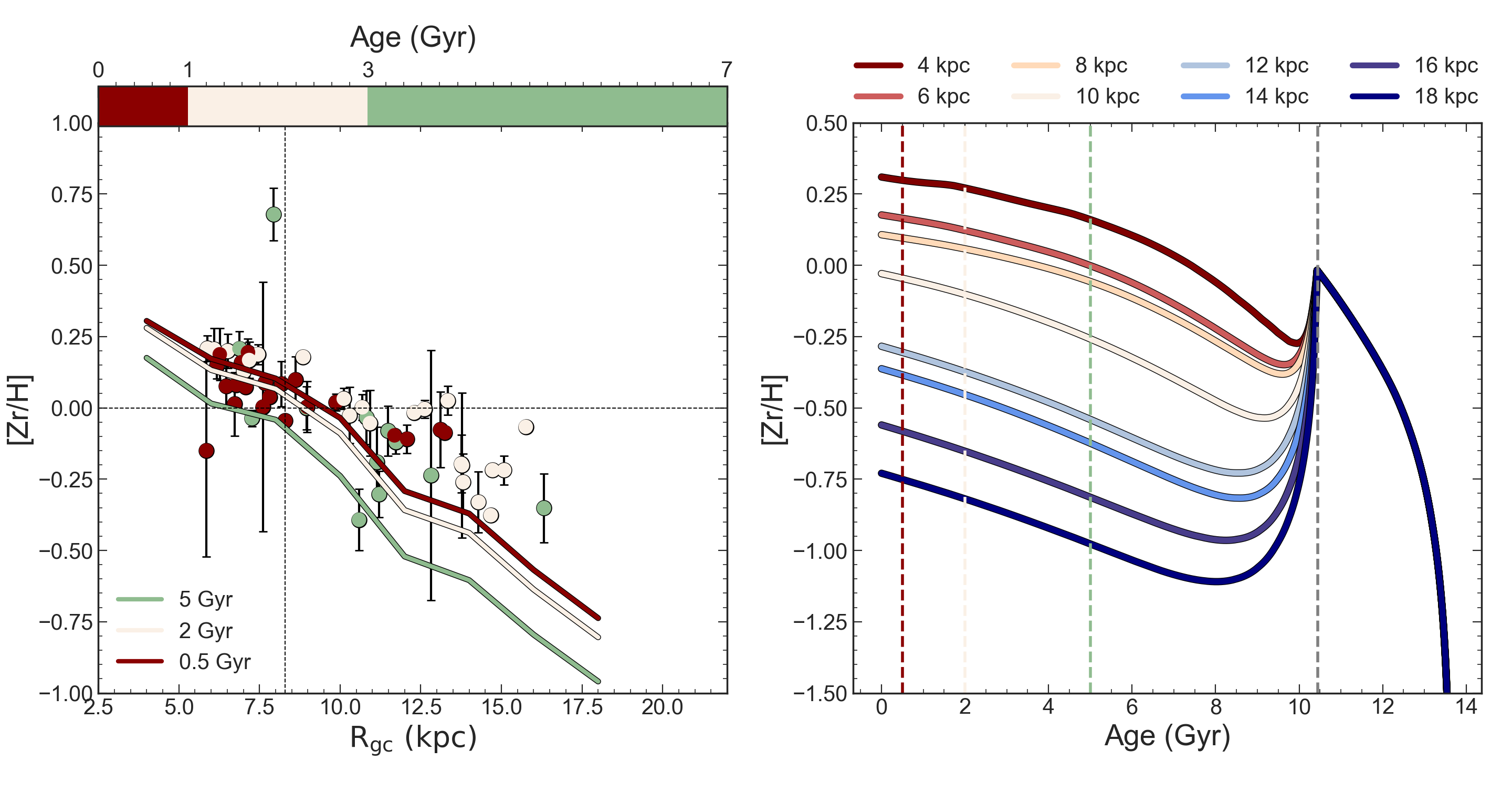}\label{fig:b}}%
    \caption{Same as Figure \ref{fig: FeH gradients} but for $\mathrm{1^{st}}$ peak s-process elements.}%
    \label{fig: gradients s-elements first peak}%
    \end{center}
\end{figure}

\begin{figure}
    \begin{center}
    \subfloat{\includegraphics[width=1\columnwidth]{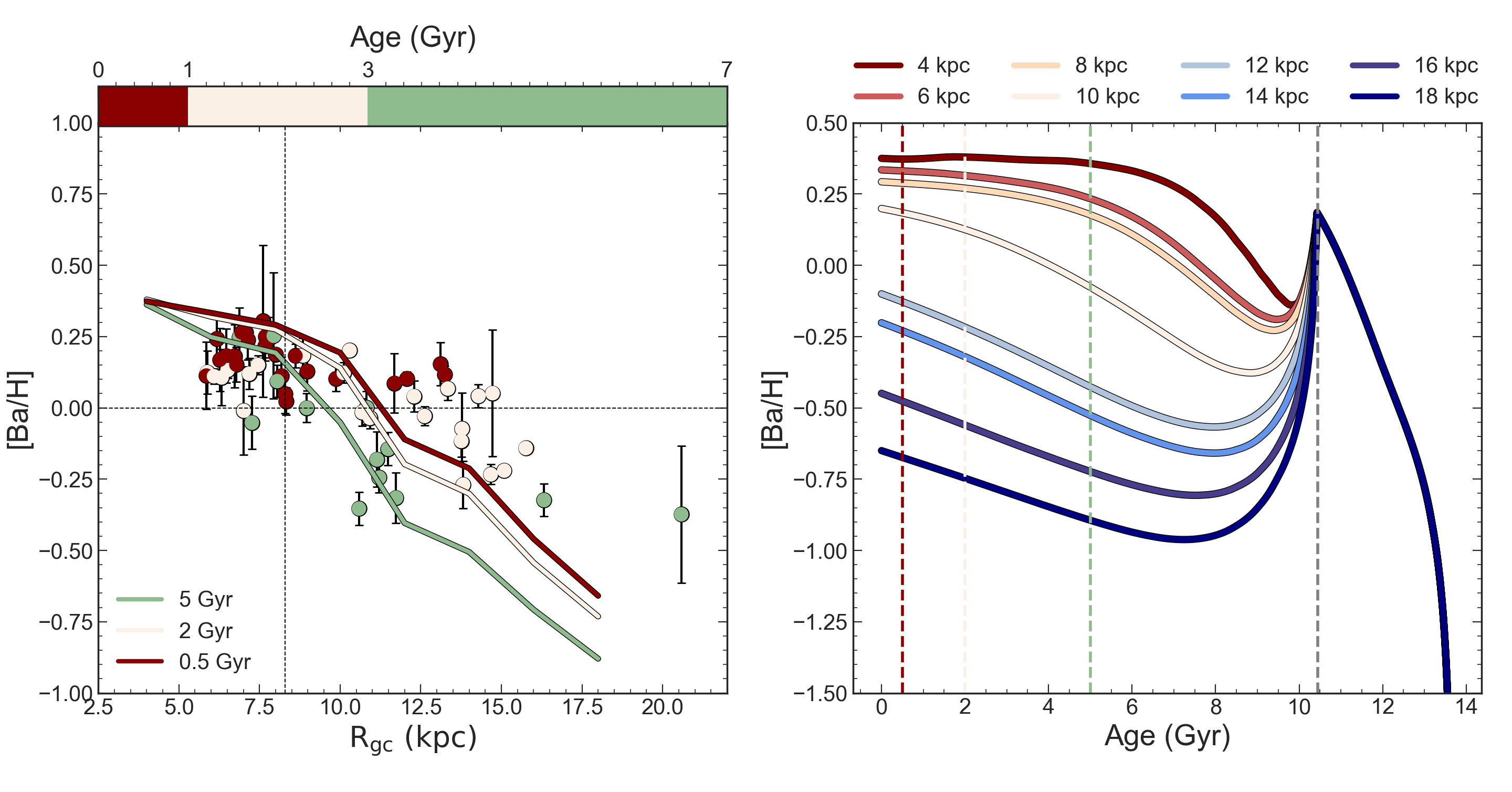}\label{fig:c}}
    \vspace{-0.3cm}
    \subfloat{\includegraphics[width=1\columnwidth]{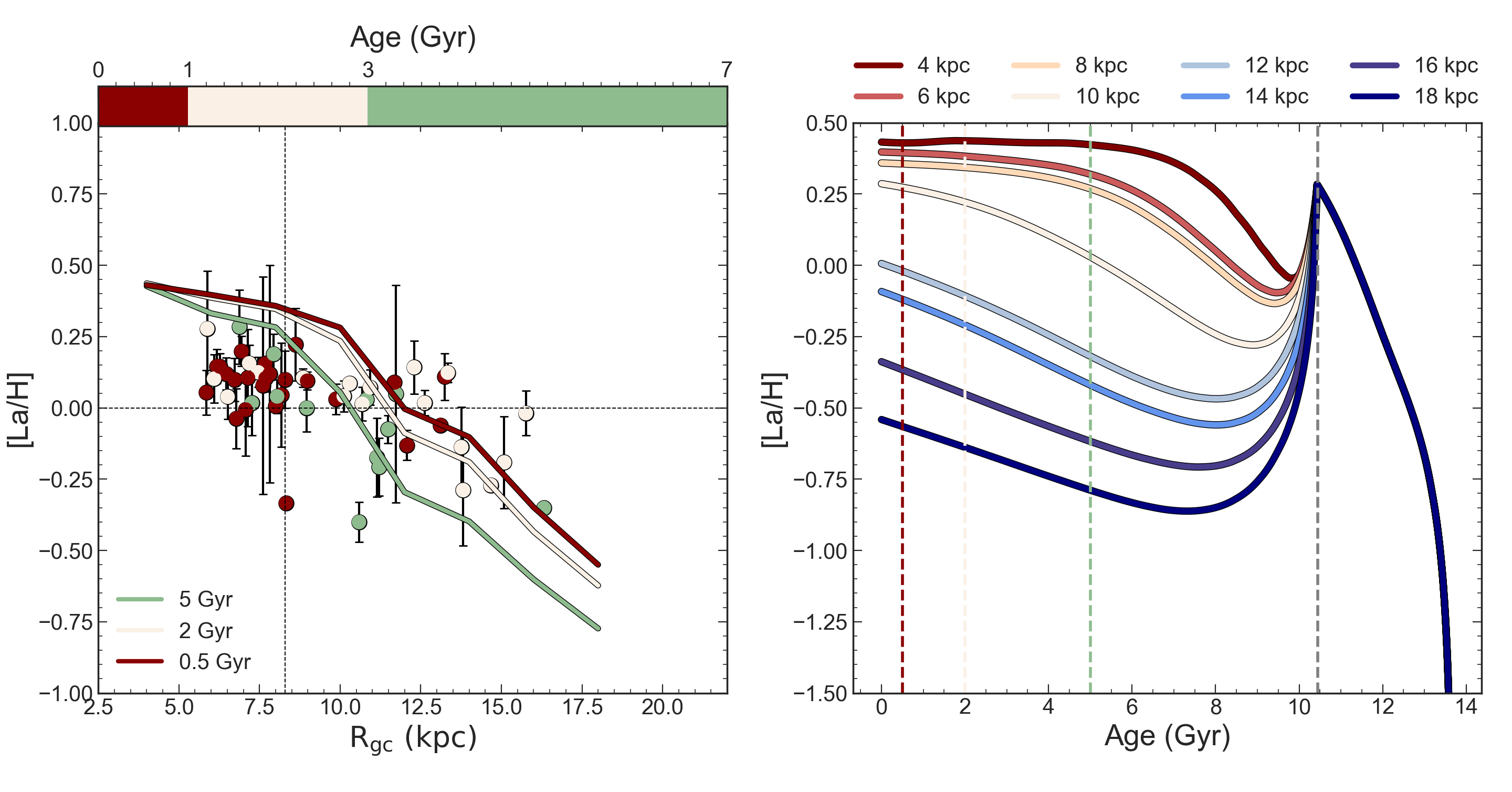}\label{fig:d}}
    \vspace{-0.3cm}
    \subfloat{\includegraphics[width=1\columnwidth]{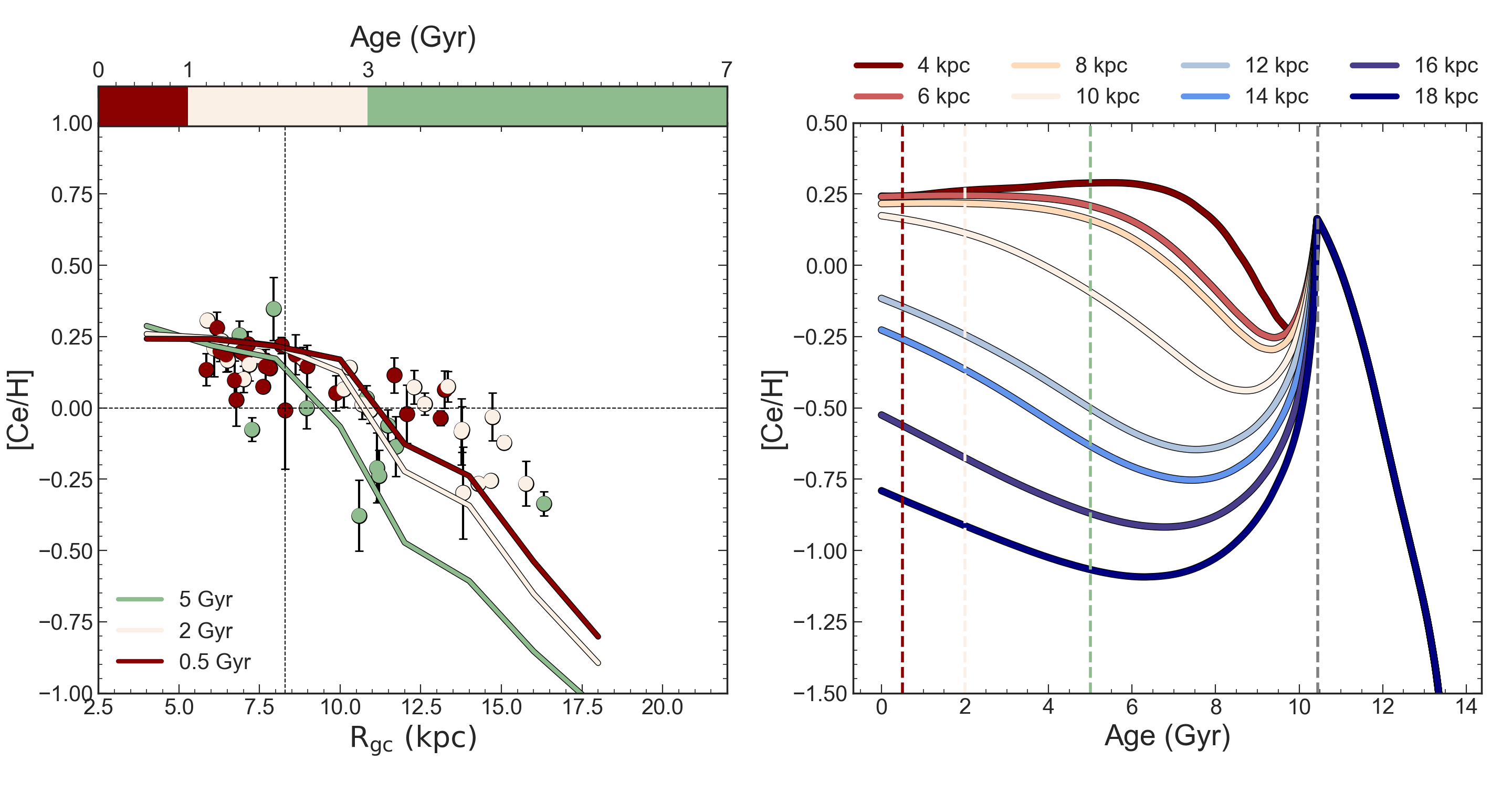}\label{fig:e}}%
    \caption{Same as Figure \ref{fig: FeH gradients} but for $\mathrm{2^{nd}}$ peak s-process elements.}%
    \label{fig: gradients s-elements second peak}%
    \end{center}
\end{figure}

In Figures \ref{fig: gradients s-elements first peak} and \ref{fig: gradients s-elements second peak}, we display the time evolution of the first and second peak s-process elements, respectively. In this case, the OC sample shows much more scatter. According to M23, the OC sample is characterized by an inverse main trend with respect to [Fe/H], with the youngest clusters being characterized by lower (or almost equal) abundances of Y, Zr, Ba, La and Ce than their older counterparts. 
%In particular, for the elements belonging to the first peak (Y and Zr) the youngest clusters seem to have a trend similar to the intermediate age ones, while for the second peak s-process elements (Ba, La and Ce) the trend appears slightly higher, at least in the outer zones.
Predictions of our model are in agreement with these trends. For Y and Zr the lines corresponding to ages of $0.5$ and $\mathrm{2\ Gyr}$ show a very similar pattern and are characterized by both the same slope and almost the same abundances at all Galactocentric distances. Similarly, in the case of Ba, La and Ce, our model predicts an almost identical flat or slightly decreasing pattern at all ages in the inner zones ($\mathrm{R_{GC}<10\ kpc}$), while the predictions diverge for higher $\mathrm{R_{GC}}$ values. The plateau observed for the s-process elements belonging to the second peak at low Galactocentric distances, is due to the effect of LIMS, which contribution reaches a maximum value faster in the inner regions than in the outer ones (see also \citealp{Casali2023}). The slopes predicted by our model for the first peak s-process elements are globally flatter with respect to those expected for the second peak elements. This is because Y and Zr are mainly produced by rotating massive stars (\citealp{Limongi2018}) and therefore on quicker timescales with respect to Ba, La and Ce which, on the other hand, have a production dominated by LIMS (\citealp{Cristallo2009, Cristallo2011, Cristallo2015}).

\begin{figure}
    \begin{center}
    \subfloat{\includegraphics[width=1\columnwidth]{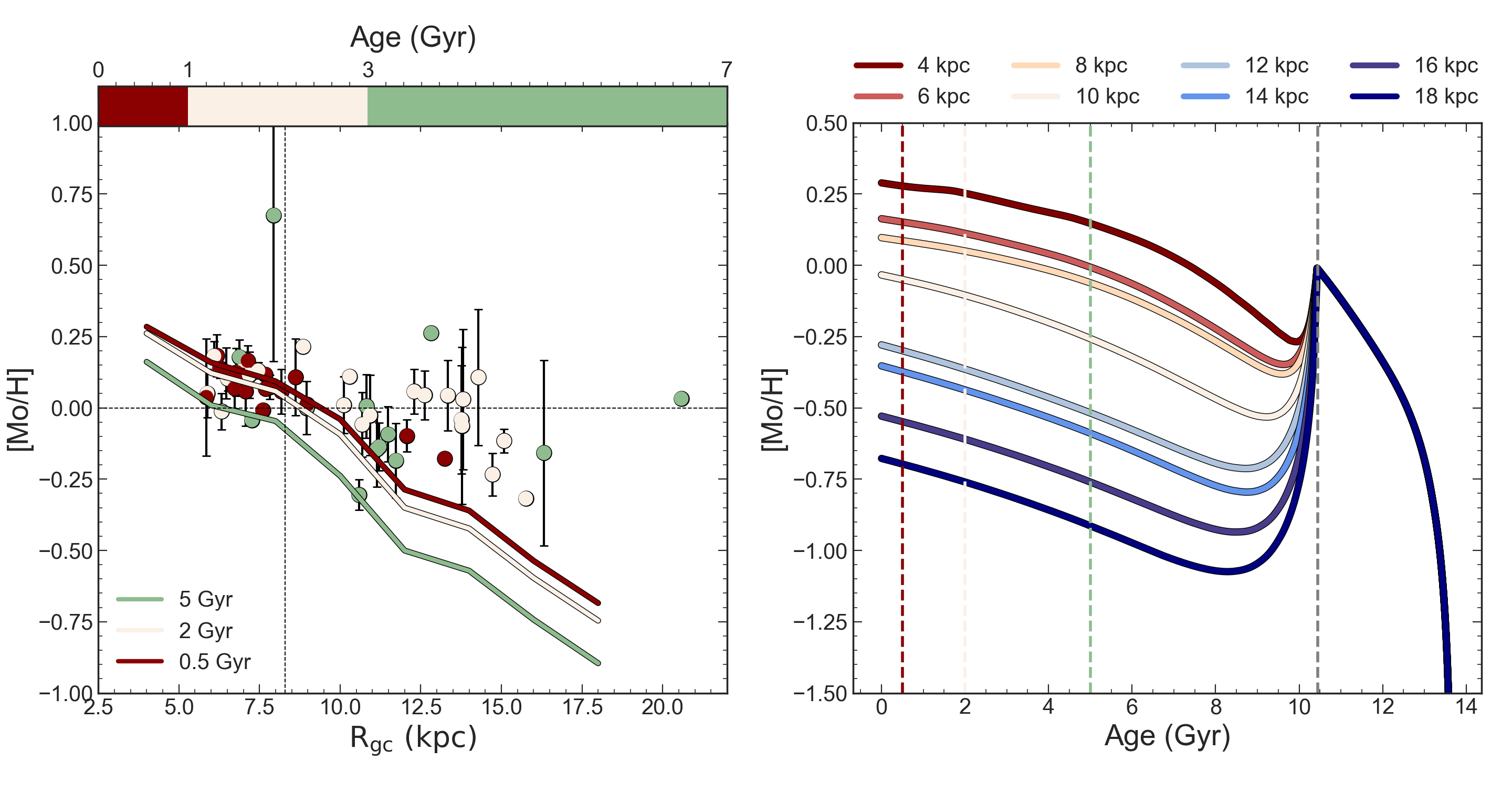}\label{fig:a}}
    \hfill
    \subfloat{\includegraphics[width=1\columnwidth]{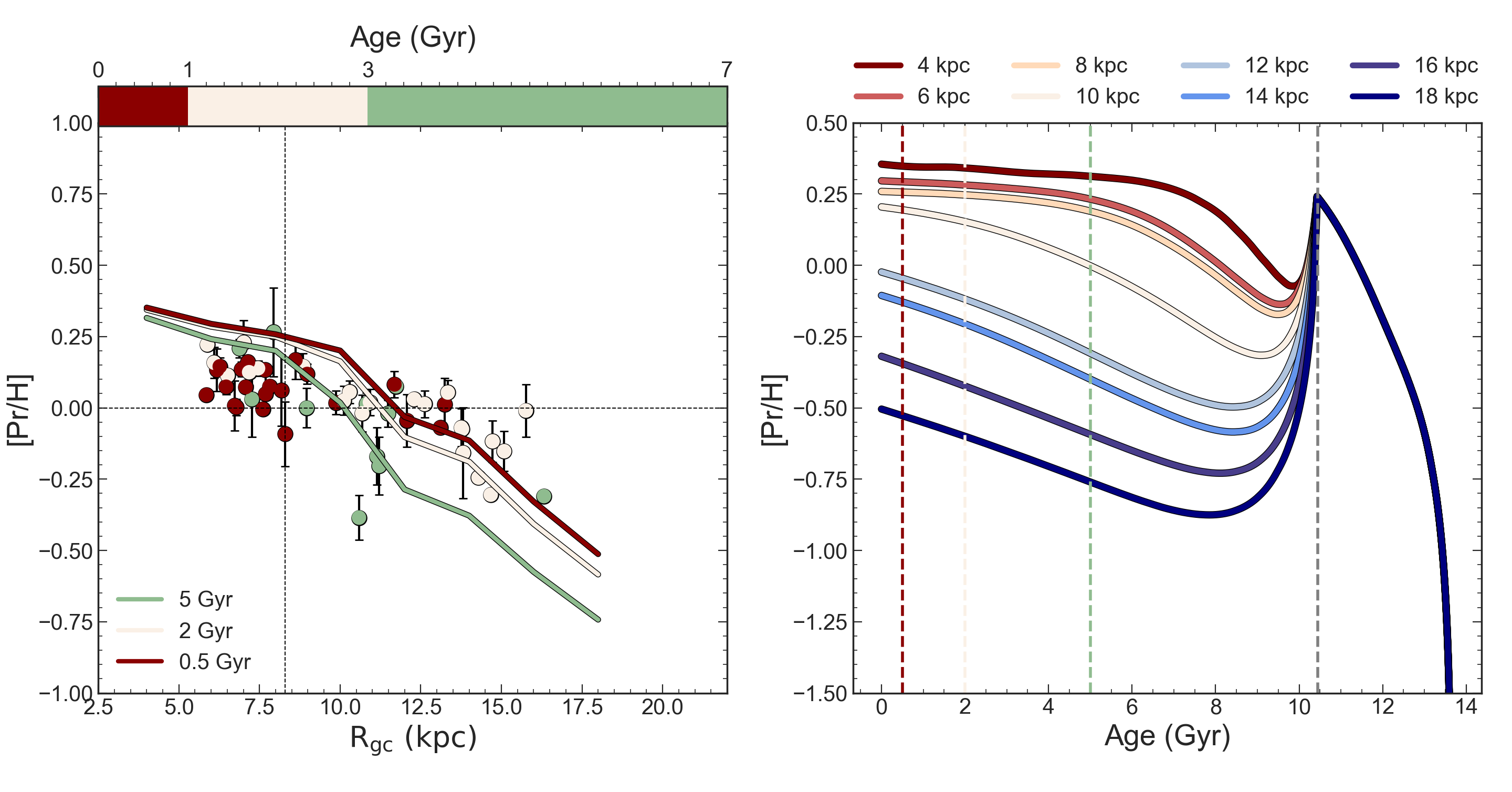}\label{fig:b}}
    \hfill
    \subfloat{\includegraphics[width=1\columnwidth]{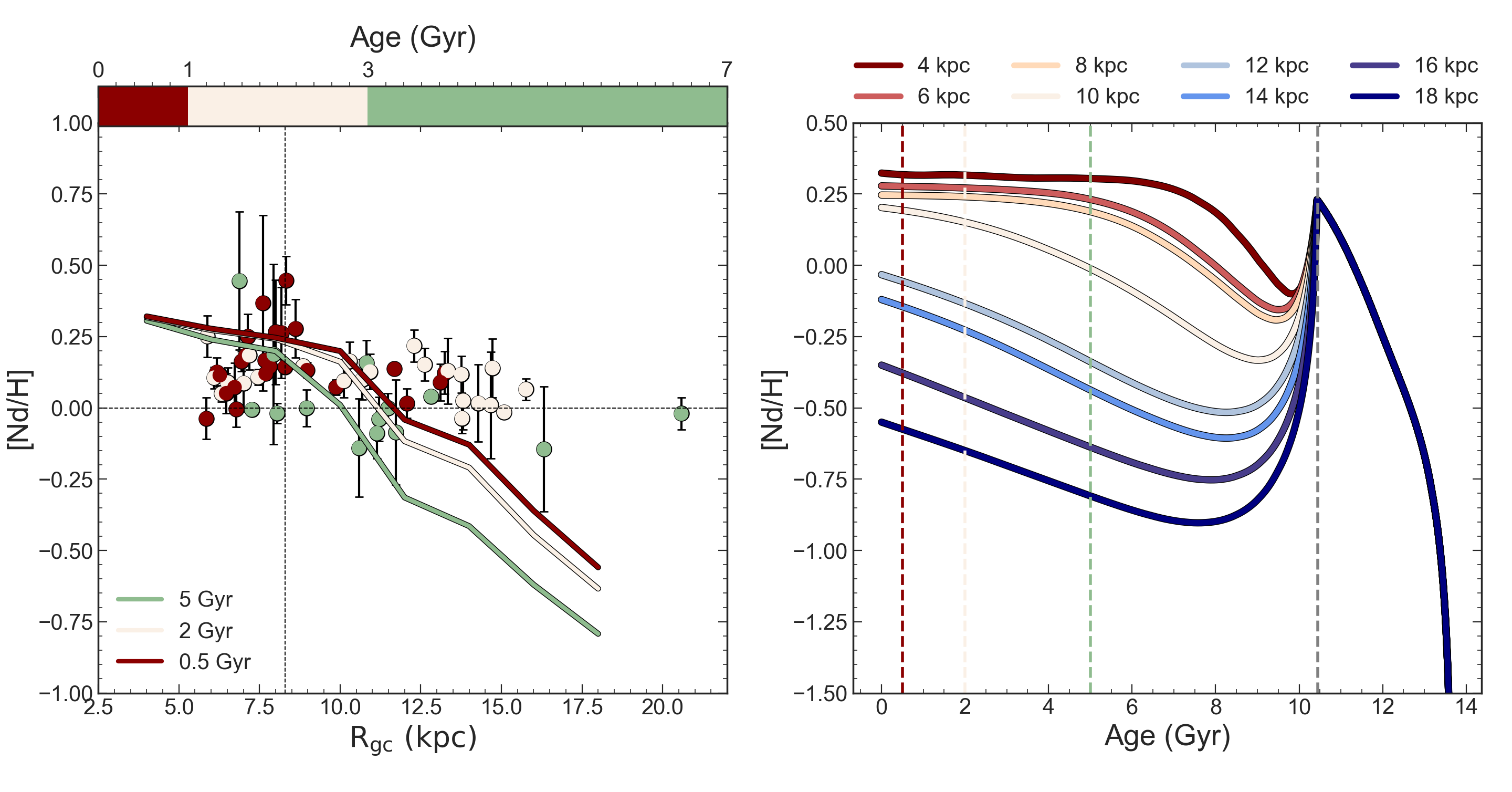}\label{fig:c}}
    \caption{Same as Figure \ref{fig: FeH gradients} but for mixed-process elements.}%
    \label{fig: gradients r-elements}%
    \end{center}
\end{figure}

The time evolution of the gradients of the other r/mixed-process elements is reported in Figure \ref{fig: gradients r-elements}. The OC sample shows a slope similar to that of Eu ($\mathrm{\sim -0.002\ dex\ kpc^{-1}}$, see M23), but, as in the case of the s-process elements, also the mixed/r-process elements are characterized by a larger scatter, in particular Mo and Nd. Unlike the [Fe/H] gradient, in the case of those elements the youngest population does not appear to be more abundant than the oldest one. As already discussed in Section \ref{sec: mixed elements}, because of the specific nucleosynthesis prescriptions adopted in this study our model underestimate the observed Mo and Nd abundances. It is possible to note that a similar plateau in the inner region predicted by our model for the s-process elements belonging to the second peak appears also for Nd and Pr. Once again, this may be due to the contribution from LIMS which reaches its maximum value faster in the inner regions.

\begin{table}
\centering
\begin{tabular}{lcccc}
\hline
    &   $\mathrm{Age=0.5\ Gyr}$   &   $\mathrm{Age=2\ Gyr}$ &   $\mathrm{Age=5\ Gyr}$\\
\hline
$\mathrm{[Fe/H]}$  & $-0.067\pm0.002$   &   $-0.069\pm0.002$   &  $-0.073\pm0.003$  \\
$\mathrm{[Eu/H]}$  &  $-0.051\pm0.003$  &   $-0.053\pm0.004$   &  $-0.059\pm0.004$  \\
\hline
$\mathrm{[Y/H]}$   &  $-0.067\pm0.004$  &   $-0.070\pm0.004$   &  $-0.077\pm0.005$  \\
$\mathrm{[Zr/H]}$  &  $-0.075\pm0.004$  &   $-0.078\pm0.004$   &  $-0.083\pm0.004$  \\
$\mathrm{[Ba/H]}$  &  $-0.078\pm0.008$  &   $-0.084\pm0.007$   &  $-0.095\pm0.006$  \\
$\mathrm{[La/H]}$  &  $-0.073\pm0.008$  &   $-0.080\pm0.007$   &  $-0.092\pm0.006$  \\
$\mathrm{[Ce/H]}$  &  $-0.077\pm0.011$  &   $-0.087\pm0.011$   &  $-0.105\pm0.008$  \\
\hline
$\mathrm{[Mo/H]}$  &  $-0.071\pm0.003$  &   $-0.073\pm0.003$   &  $-0.077\pm0.004$   \\
$\mathrm{[Pr/H]}$  &  $-0.078\pm0.012$  &   $-0.089\pm0.012$   &  $-0.109\pm0.009$  \\
$\mathrm{[Nd/H]}$  &  -$0.058\pm0.004$  &   $-0.062\pm0.004$   &  $-0.070\pm0.005$  \\
\hline
\end{tabular}%
\caption{\label{tab: gradients with ages} Slopes of the [El/H] gradients predicted by our model at $\mathrm{Age=0.5}$, $2$ and $\mathrm{5\ Gyr}$.}
\end{table}

\section{Summary and conclusions}
\label{sec: conlusion}

In this paper we studied the origin of neutron capture elements in the MW by taking advantage of the large sample of OCs from the \textit{Gaia}-ESO DR6. To this aim we adopted the revised two-infall model (\citealp{Palla2020}; see also \citealp{Spitoni2019}). We investigated the abundance patterns and the radial gradients of 5 s-process (Y, Zr, Ba, La and Ce) and 4 mixed/r-process elements (Eu, Mo, Nd and Pr). In order to do that, we adopted the following nucleosynthesis prescriptions: s-process material is produced by i) rotating massive stars ($\mathrm{M>13\ M_\odot}$) with yields from \citet{Limongi2018} with three different initial rotational velocities (0, 150, and 300 km/s) and by ii) LIMS ($\mathrm{1\leq M/M_\odot\leq 8}$) with yields from the FRUITY data base (\citealp{Cristallo2009, Cristallo2011, Cristallo2015}) in the $\mathrm{1-6\ M_\odot}$ range, arbitrarily extrapolated up to $\mathrm{8\ M_\odot}$. R-process material is produced by both a prompt and a delayed source, namely i) MR-SNe which are supposed to be 20\% of all massive stars with initial mass between $\mathrm{10-25\ M_\odot}$ with yields from \citet{Nishimura2017} (model L0.75) and ii) MNS with a DTD from \citet{Simonetti2019} with $\mathrm{\beta=-0.9}$ with yields prescriptions from \citet{Molero2021_2} best model. Our conclusions for the abundances patterns of [El/Fe] vs. [Fe/H] can be summarized as follows:

\begin{itemize}
    \item The [Eu/Fe] vs. [Fe/H] abundance pattern is well reproduced if both a quick source and a delayed one act as r-process producers. This is a well known result in chemical evolution. However, here we stress how with the assumed prescriptions the quick source completely dominates the production of Eu, in agreement with the recent work of \citet{vanderswaelmen2023} according to which there is no need for an additional delayed source at least to reproduce the abundance pattern in the thin disc. However, since MNS are the only observed source of neutron capture elements up to now, they cannot be excluded from chemical evolution models computations.
    \item The s-process elements abundances patterns is not reproduced if one considers only production from typical s-process astrophysical sources as rotating massive stars and LIMS. Rotation increases the production of s-process material, especially at low metallicities and for elements belonging to the first s-process peak, but the r-process component must also be taken into account.
    \item When the contribution from MNS and MR-SNe to the production of the r-process component of the s-process elements is added, MR-SNe dominate at low metallicities and it is no longer possible to appreciate differences between different rotational velocities for massive stars. The s-process abundance pattern of the OC sample is well reproduced. The rise in the outer disc data as well as the peak followed by the decrease in the local and inner-disc are reproduced by our model at $\mathrm{R_{GC}=12}$, 8 and $\mathrm{6\ kpc}$. Only for [Y/Fe] and [Zr/Fe] vs. [Fe/H] in the outer region our model predicts a decrease rather than the observed increase, due to too low yields of these elements from LIMS and too high ones from MR-SNe.
    \item The picture for the mixed/r-process elements is more complex. A good agreement with the relevant data is obtained for Mo, even if the model still slightly underestimates the observations, most probably because of the lack of an additional contribution from neutrino-driven SNe. For Nd and Pr, we disagree with \citet{vanderswaelmen2023}, who claim that Nd is characterized by a significant s-process contribution, whereas Pr by a lower one. On the contrary, our model shows a higher production from LIMS of Pr than of Nd. Also in this case the model underproduces the [Nd/Fe] vs. [Fe/H] abundance pattern, while nicely reproduces the [Pr/Fe] vs. [Fe/H] one.
\end{itemize}

As for the abundance gradients, we first compared predictions of our model for the present day radial gradients of [Fe/H] and [Eu/H] with those traced by GES DR6 OCs with $\mathrm{Age<3\ Gyr}$ and Cepheids from \citet{Luck2011} and \citet{Genovali2015}. Then we discussed the evolution with time of the abundance gradients of all the neutron capture elements studied in this work in comparison with the full OC sample. Our conclusions can be summarized as follows:

\begin{itemize}
    \item The present day slope of the [Fe/H] gradient predicted by our model is $\mathrm{-0.067\pm 0.003\ dex\ kpc^{-1}}$, slightly steeper both with respect to that of the restricted OC sample and to the one of the Cepheid sample, yet it agrees with other recent slopes of the [Fe/H] gradient from OC samples (\citealp{Carrera2019,Donor2020,Zhang2021,Spina2021,spina2022}).
    \item The flat slope observed in the OC sample for [Eu/H] is not reproduced by the model in which Eu is produced by a quick and a delayed source (MR-SNe + MNS with a DTD). Models with no delayed source (only with MR-SNe or only with MNS with a constant and short delay time for merging) do not improve the fit to the data. We discussed the possibility of flattening the predicted [Eu/H] gradient by adopting a constant SF efficiency. However, we are not inclined to relax the assumption of a variable SF efficiency, since it has already been proved by many authors (e.g. \citealp{colavitti2009, spitoni2015, Grisoni2018, Palla2020}) that the inside-out scenario by itself is not able to explain the abundance patterns at different Galactocentric distances and the abundance gradients for several elements, as well as the gradient of the SFR and gas density along the thin disc (see \citealp{Palla2020}). A reasonable explanation for the discrepancy between model results and observations could thus be that clusters with intermediate age ($\mathrm{1 \leq Age \leq 3\ Gyr}$) are affected by radial migration. In favor of this hypothesis, predictions of our model are much more in agreement with the slopes observed in the inner-disc rather than with the outer ones and a better agreement is also obtained with the radial gradients of the Cepheid sample.
    \item Regarding the time evolution of the [Fe/H] gradient, results of our model for $\mathrm{Age=0.5}$, 2 and $\mathrm{5\ Gyr}$ are in agreement with the observed trend if the gradient of the youngest population is computed by removing all giant stars with log$g$$\mathrm{<2.5}$ (see M23 for details). In particular, a really limited time evolution of the [Fe/H] gradient between the considered ages should be expected.
    \item Also for the [Eu/H] gradients a limited evolution with time is predicted by our model, in agreement with the observations. However, as already seen in the case of the present day gradient also at different ages we obtain slopes which are too steep with respect to the observations. 
    \item Predictions of our model for the radial [Y/H] and [Zr/H] gradients show a very similar pattern for $\mathrm{Age=0.5}$ and $\mathrm{2\ Gyr}$, in agreement with the OC sample. Also in the case of Ba, La and Ce the model predicts an almost identical flat or slightly decreasing pattern for all ages in the inner zone, as observed in the OC sample.
    \item As for the abundance patterns, also for the radial gradients much more uncertainty is present in reproducing the trend of the other mixed/r-process elements. Due to the adopted nucleosynthesis prescriptions, we underestimate the trends for Mo and Nd, and always produce steeper gradients with respect to the observed ones.
    \item For all the elements belonging to the second s-process peak as well as for Nd and Pr, our model produces a plateau for low Galactocentric distances at all the considered ages. This is most probably due to the effect of LIMS, whose production of those elements reaches an equilibrium value before that in the outer regions, as a consequence of a faster SF.
    \end{itemize}

\section*{Acknowledgements}

We thank the anonymous referee for the useful comments which improved the manuscript. M. Molero and F. Matteucci thank INAF for the 1.05.12.06.05 Theory Grant- Galactic archaeology with radioactive and stable nuclei. M. Molero thanks Federico Rizzuti for the useful discussions. M. Palla acknowledges funding support from ERC starting grant 851622 DustOrigin. C. Viscasillas acknowledges funding from the Lithuanian Science Council (LMTLT, grant No. P-MIP-23-24). E. Spitoni  received funding from the European Union’s Horizon 2020 research and innovation program under SPACE-H2020 grant agreement number 101004214 (EXPLORE project). 
%%%%%%%%%%%%%%%%%%%%%%%%%%%%%%%%%%%%%%%%%%%%%%%%%%

\section*{Data Availability}

The data underlying this article will be shared upon request.



%%%%%%%%%%%%%%%%%%%%%%%%%%%%%%%%%%%%%%%%%%%%%%%%%%

%%%%%%%%%%%%%%%%% APPENDICES %%%%%%%%%%%%%%%%%%%%%

%\appendix



%%%%%%%%%%%%%%%%%%%%%%%%%%%%%%%%%%%%%%%%%%%%%%%%%%

% Don't change these lines
\bsp	% typesetting comment
\label{lastpage}
\end{document}